\documentclass[11pt,a4paper,english]{article}
\usepackage[latin9]{inputenc}
\usepackage{float}
\usepackage{amsmath}
\usepackage{graphicx}
\usepackage{esint}
\usepackage{array}
\usepackage{float}
\usepackage{amsmath}
\usepackage{stackrel}
\usepackage[version=4]{mhchem}
\usepackage[running]{lineno}

\usepackage{listings}
\usepackage{listings}
\usepackage{color}

\definecolor{dkgreen}{rgb}{0,0.6,0}
\definecolor{gray}{rgb}{0.5,0.5,0.5}
\definecolor{mauve}{rgb}{0.58,0,0.84}

\makeatletter

\pdfoutput=1 

\usepackage{jinstpub}

\title{\boldmath Numerical study of effects of electrode parameters and image charge on the electric field configuration of RPCs}

\author[a,c,1]{Tanay Dey,\note{Corresponding author.}}
\author[a,b]{Supratik Mukhopadhyay,}
\author[c]{Subhasis Chattopadhyay}

\affiliation[a]{Homi Bhabha National Institute,\\Mumbai, India}
\affiliation[b]{Saha Institute of Nuclear Physics,\\Kolkata,India}
\affiliation[c]{Variable Energy Cyclotron centre,\\Kolkata, India}

\emailAdd{tanay.jop@gmail.com}

\abstract{The working of a resistive plate chamber depends on the electric field applied inside the gas gap. The strength of this electric field inside the gas gap depends on parameters like electrode thickness, permittivity, gas gap, among others. The applied electric field can get significantly modified by space charge effect. This can be a major concern while working in high rate particle physics experiments. Accumulation of space charge generated due to consecutive avalanches can distort the applied field. During this investigation, we have observed that the strength of this perturbation in electric field is a function of the amount of charge, electrode thickness, permittivity, and gas gap. 
	In this paper, we have studied the dependence of the applied field on parameters linked to  RPC fabrication, mentioned above. In addition, we have used the method of image to calculate the additional polarisation field due to space charge for a basic five layers of the geometry of an RPC.  The variation of the perturbation in electric field with the same three parameters electrode thickness, permittivity, gas gap has also been discussed. }


\arxivnumber{} 

\collaboration[c]{}

\@ifundefined{showcaptionsetup}{}{%
	\PassOptionsToPackage{caption=false}{subfig}}
\usepackage{subfig}
\makeatother

\usepackage{babel}
\begin{document}
	\maketitle \flushbottom
	\section{Introduction}
	A resistive plate chamber (RPC)\cite{cardeli-1,cardeli-2} is a low-cost gaseous particle detector,
	which has simple planar geometry. High efficiency, good time resolution,
	and the possibility of being tailored to any shape make it very useful in HEP experiments like INO, CMS, ALICE, CBM etc\cite{Goswami_2017,Kumari_2020,Collaboration_2012,MONDAL2021166042}. The basic detector physics parameters like ionization, multiplication, signal induction, etc., are functions of temperature, pressure, and electric field.
	So, a meticulous field measurement is always necessary. It is known that the direct measurement of the electric field inside the RPC is a difficult task to achieve. However, an indirect measurement of the electric field in the gas discharge process using stark shift exists in the literature \cite{Cvetanovi__2015}. As a result, it becomes necessary to rely on the analytical or numerical method to estimate the electric field inside RPC. An analytical approach of field calculation using the surface charge method (SCM) inside an RPC can be found in \cite{AMMOSOV1997217}. However, in order to solve realistic, complex problems, it is preferable to use a numerical approach due to having limitations in the analytical approach.
	In a single gap RPC, other than three primary layers (two electrodes and a gas gap), there are other dielectric layers and geometric non-uniformities such as side and button spacers. The electric field in such geometries with distinctly three-dimensional features can be obtained using numerical solvers such as neBEM and COMSOL. \cite{MAJUMDAR2008346,MAJUMDAR2009719,comsol}.
	
	RPC offers three basic operation modes, which are a) avalanche mode,
	b) saturated avalanche mode, and c) streamer mode \cite{MOSHAII2012S168,Cardarelli1996AvalancheAS}. The charge accumulation near the electrode is high due to successive avalanches in high rate experiments.
	As an effect, the distortion in the applied field due to the space charge effect is also significant \cite{Lippmann_1}. This phenomenon slows down the growth of avalanches generated from subsequent primaries, which reduces the particle detection efficiency \cite{rpc-book}. Therefore, limiting the average charge is an efficient way to work at a high rate\cite{Paolozzi:2012pt}. Indeed, in high rate experiments, it is preferable to work
	in avalanche mode or low gain mode rather than the other two modes since gain, and consequently the generation of space charge, is larger in these other modes.
	
\begin{figure}
		\center{\includegraphics[scale=0.36]{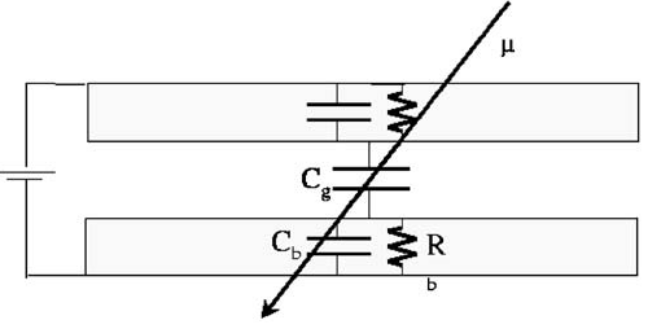}
			
		}
		
		\caption{\label{fig:equivalentRPC}Equivalent circuit of an RPC. \cite{ABBRESCIA20047}}
	\end{figure}

	The equivalent circuit of an RPC system has been shown in figure \ref{fig:equivalentRPC}. The gas gap of the RPC  is represented as the capacitor $C_{g}$ and the resistive electrodes are represented as a parallel combination of capacitor $C_{b}$ and resistance R. The capacitor $C_{g}$ can be partially discharged when an ionizing particle passes through the gas gap and the effective voltage across the gas gap reduces proportionally to the generated avalanche charge $q_{av}$. Then, the voltage across $C_{g}$ restore with the help of an external power supply. The behavior of the charging up process is exponential, and the RC time constant $\tau_{g}$ of an RPC can be written as \cite{Gonzalez-Diaz:2006qno,ABBRESCIA20047}: 
	\begin{equation}\label{eq:RC_FORMULA}
	\tau_{g}=2R_b(2C_b+C_g)\\
	=\rho\epsilon_{0}(\epsilon_{r}+\frac{d}{g}),
	\end{equation} 
	where $\rho,\epsilon_{r},d$ are resistivity, relative permittivity, the thickness of electrodes respectively, and $g$, $\epsilon_{0}$ is the gas gap and free space permittivity. The bulk resistivity ($\rho$) of bakelite electrodes may vary between  $\backsim10^9-10^{11}\Omega\mbox{-}m$. Now let the gas gap (g) and electrode thickness (d) is 2mm and relative permittivity ($\epsilon_r$) of bakelite is 5. Then, for $\rho\approx2\times10^{10}\Omega\mbox{-}m$ the typical value of the $\tau_g$ can be calculated using equation \ref{eq:RC_FORMULA} as follows:
	\begin{equation}
	\tau_g=(2\times 10^{10})\times(8.85\times10^{-12})\times(5+1)=1.05 sec.
	\end{equation}
	 A low $\tau_{g}$ of electrodes can serve a good detection
	rate. Hence, a search for optimized electrodes based on the parameters
	$\rho,\epsilon_{r,}d,g$ is necessary \cite{Aielli_2016,Carboni:2003my}. A detailed simulation of the rate capability may help to optimize the same parameters. 
	The accuracy of such simulation depends on the precise calculation of the dynamic space-charge electric field along with the two polarisation fields of electrodes a) polarisation due to space charge field and b) DC
	polarisation (applied field). It is known that dielectric materials show a polarisation effect in the presence of the electric field. The bound charges in each molecule of the dielectric electrodes are influenced by the applied field and execute the perturbed motions, which distort the molecular charge density. Without an external field, the average dipole moment of the molecules of any dielectric material is usually zero. In the presence of an external field,  there would be a net dipole moment inside the dielectrics due to dipole alignment towards the field. Thus, the electric polarisation $\vec{P}$ is produced in the medium (net dipole moment per unit volume). The effect of polarisation yields net volume and surface-bound charges. These bound charges, along with other free charges inside the dielectric electrode, are responsible for the electric field inside the gas gap of the RPC. These polarisation fields can be calculated by solving the Poisson equation with proper boundary conditions or using the image method. A
	Poisson equation solving approach can be found in \cite{HEUBRANDTNER,Lippmann:2003ar,Lippmann_1}. Alternatively the polarisation field due to space charge can be calculated using method of the image for such a simple geometric configuration, which  is a relatively easier and faster way. The basic geometry
	of an RPC consists of five layers of dielectrics as shown in
	 figure \ref{fig:RPC_image_charge_formation}.
	The formation of the image due to three layers of the dielectric can be found in the section 21. of chapter 6 of \cite{Weber-W}, where the charge resides inside the middle dielectric material
	 and the dielectric constant of outermost dielectrics is considered the same. The geometrical configuration used in the section 21. of chapter 6 of \cite{Weber-W} is nearly similar to that of an RPC. However, in that work the point charge location is not general, and the termination condition of infinite series of reflections for a general position of charge is also not discussed. A more detailed formation of infinite series of images of a point charge positioned in front of three dielectric layers has been discussed in \cite{Jomaa1983ElectricFD,multi-layer}(see figure \ref{fig:threelayer}). 
	 
	 In this paper, in section \ref{sec:section2} we have discussed the variation of the applied DC field inside the gas-gap with the three RPC parameters, i.e., $\epsilon_r d$ and $g$. In section \ref{sec:section3}, we have first discussed the method of image for metal electrode and then generalized the method for an RPC of five dielectric layers by considering that each dielectric electrode is different from the other. In this study, the point charge can be located anywhere inside the middle dielectric layer (gas-gap). We also gave an instance of the
	 calculation of the 3D space-charge
	 field along with its polarisation effect on the electrodes of an RPC in section \ref{sec:Section4_avalancheCharge} for a specific avalanche charge distribution using the model described in \cite{Dey_2020} and compared the results with the results from neBEM field solver and other models from the literature. We have also discussed the behavior of the electrode polarisation field with important detector parameters $\epsilon_{r,}d$ and $g$.
	
	\section{Variation of applied field inside the RPC due to different electrode parameters and gas gap} \label{sec:section2}
	A numerical model of an RPC having sides 15 cm $\times$15 cm and 10 microns thick of graphite layer has been designed in Garfield++. A DC voltage of $\pm$4.5KV is applied over the two graphite layers. However, the parameters electrode thickness $d$, gas gap $g$, and relative permittivity $\epsilon_{r}$ are varied systematically, and the electric field is calculated for different configurations using the neBEM solver. This study has been divided into three cases where at a time, we fixed two parameters and varied the third one.\\
	Case 1:  $d$ and $g$ kept fixed and varied $\epsilon_{r}$.\\
	Case 2:  $\epsilon_{r}$ and $g$ kept fixed and varied $d$.\\
	Case 3:  $\epsilon_{r}$ and $d$ kept fixed and varied $g$.
	\subsection{Case 1, variation of $\epsilon_{r}$}
\begin{figure}
		\center\subfloat[\label{electricField_diff_epr}]{\includegraphics[scale=0.36]{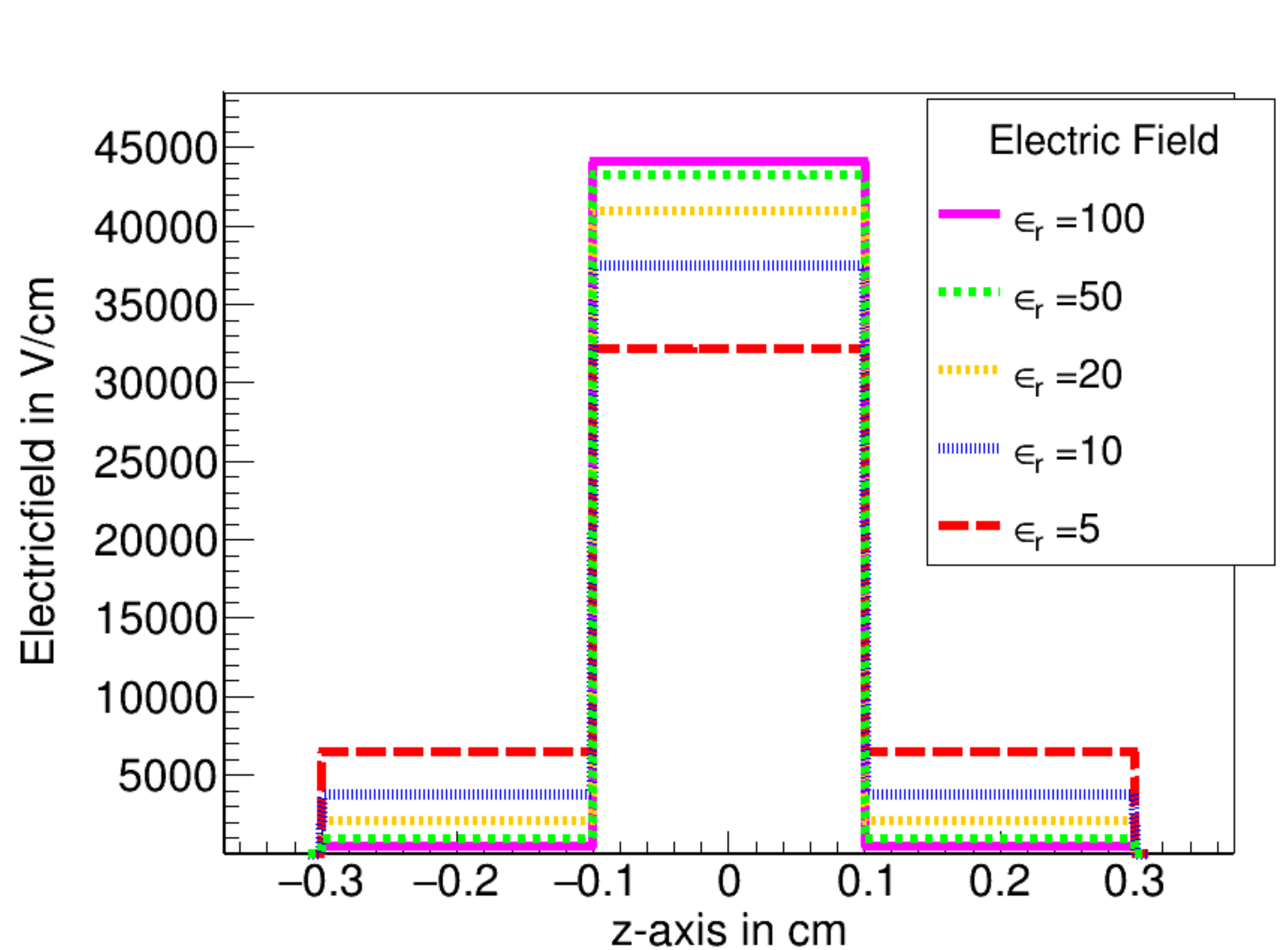}
			
		}
	
	    \subfloat[\label{pot_diff_epr}]{\includegraphics[scale=0.36]{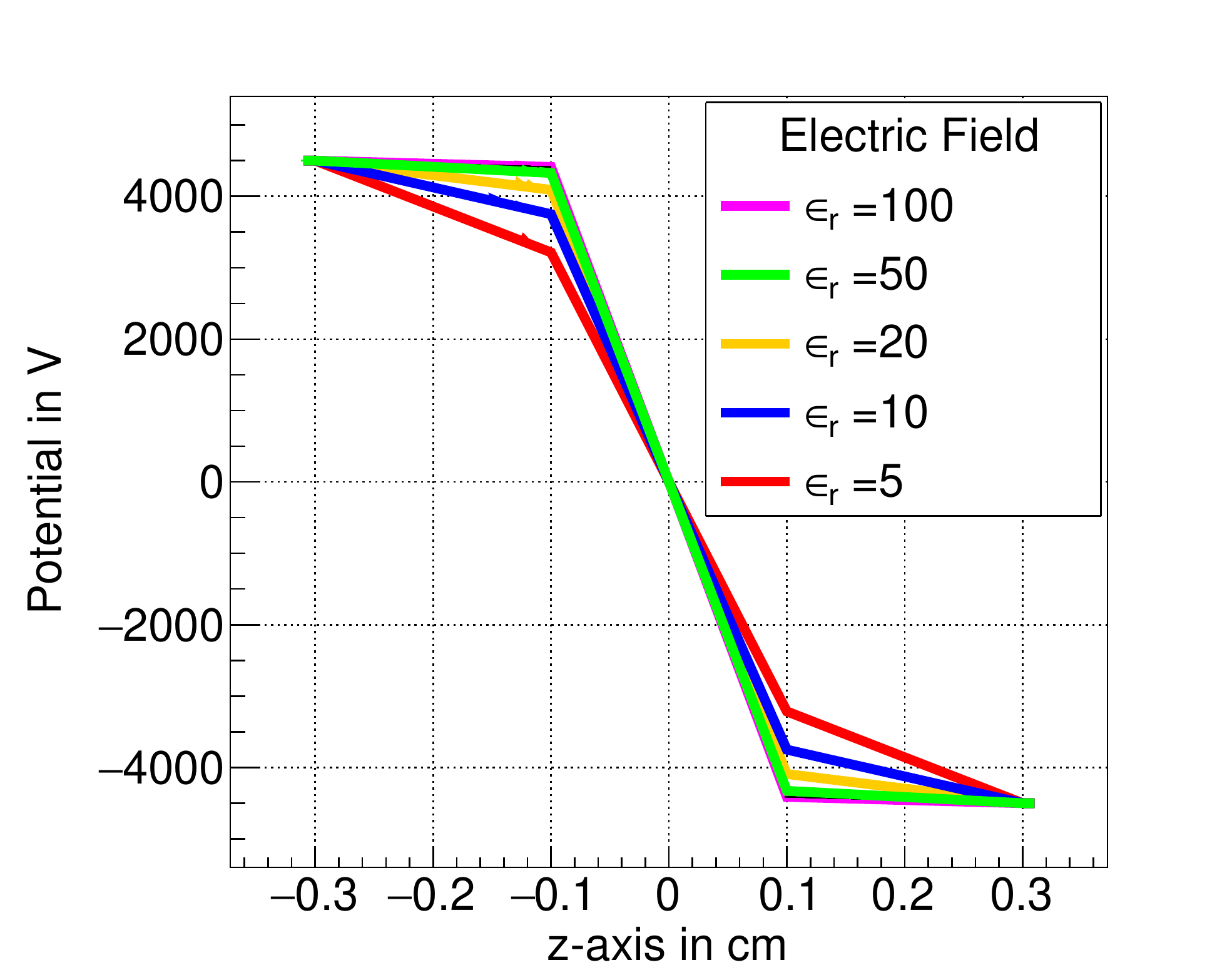}}\subfloat[\label{E_vs_epr}]{\includegraphics[scale=0.167]{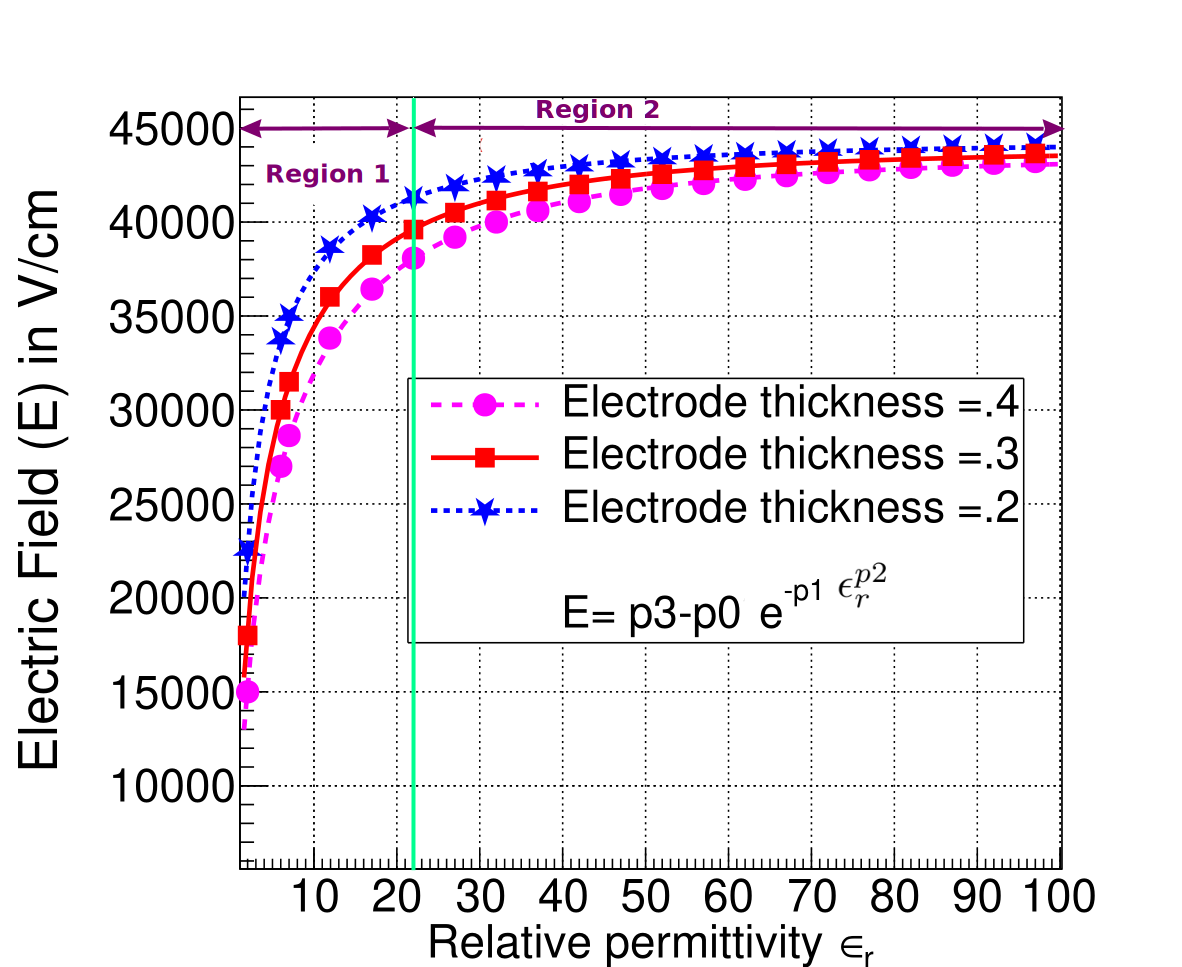}}
		
		\caption{\label{fig:electricField_diff_epsilon} (a) Applied electric field along the z-axis for different $\epsilon_{r}$ of electrodes. (b) Potential inside the RPC along the z-axis for different $\epsilon_{r}$. (c) Variation of electric field with $\epsilon_{r}$ at the middle of the gas gap. }
	\end{figure}

	It is known that the field inside the gas gap of the RPC is nearly constant at a fixed applied voltage. However, the field can be different for different electrode materials. The magnitude of the electric field and potential along the z-axis (perpendicular to the parallel plates) inside the RPC is shown in figures \ref{electricField_diff_epr} and \ref{pot_diff_epr} for different relative permittivity $\epsilon_{r}$ of electrodes. In figure \ref{electricField_diff_epr} the field is gradually increasing inside the gas gap (-0.1cm to 0.1cm) if we increase the value of permittivity $\epsilon_{r}$ but is gradually decreasing inside the dielectric electrodes (-0.3cm to -0.1cm and 0.1cm to 0.3cm) with the same $\epsilon_{r}$. Eventually, the slope of the variation of potential with z position is also dissimilar with different $\epsilon_{r}$ in all regions, which is shown in figure \ref{pot_diff_epr}. The variation of the electric field with $\epsilon_{r}$ is visible more prominently in figure \ref{E_vs_epr}, where the electric field is calculated at the center of the detector. It is clear from the figure  \ref{E_vs_epr} that the field initially grows and then saturates as $\epsilon_{r}$ is increased. We can roughly divide figure \ref{E_vs_epr} based on electric field variations and data points in two regions, a) region 1 and b) region 2. Region 1 is from $\epsilon_{r}$=2 to 22 and the region 2 is from $\epsilon_{r}$=22 to 97. The variation of the electric field with $\epsilon_{r}$ is higher than region 2. Therefore, region 2 can be considered as the saturated region. Now, the maximum percentage of change in the field on changing $\epsilon{_r}$ in both regions 1 and 2 are tabulated below (Table \ref{table_diff_epsilon}) for three different thicknesses.
From Table \ref{table_diff_epsilon} it can be said that the percentage of change in the field is consistently falling off with the thickness of electrodes (d) in both regions 1 and 2. Hence, the dependence
of the electric field on  $\epsilon_{r}$ is low for the smaller thickness of the electrodes. 
	\begin{table}[H]
		\center%
		\begin{tabular}{|c|c|c|}
			\hline 
			d (cm) & change of field in region 1 (\%)  & change of field in region 2 (\%)\tabularnewline
			\hline 
			\hline 
			0.4 & 153.8 & 13.5\tabularnewline
			\hline 
			0.3 & 120 & 10.22\tabularnewline
			\hline 
			0.2 & 83.33 & 6.88\tabularnewline
			\hline 
		\end{tabular}
		
		\caption{\label{table_diff_epsilon}Percentage of change in electric field for different thickness of electrodes.}
		
	\end{table}
	The data points of figure \ref{E_vs_epr} are fitted with the equation: 
	\begin{equation}\label{eqn:2.1}
	f(\epsilon_{r})=p3-p0\, Exp(-p1\,\epsilon_{r}^{p2}),
	\end{equation}
	to understand the functional behavior of the variation of the electric field with $\epsilon_{r}$ at a fixed electrode thickness and to allow quick interpolation at intermediate values of $\epsilon_{r}$. The fit results are tabulated below in Table \ref{table2_fitVAlue}.  From the dimensional analysis, it is clear that the dimension of the parameters p3 and p0 should have the dimension of the electric field, where p0 is normalisation constant, and p3 can be taken as the offset. Since the term inside the exponential should be dimensionless, therefore the parameter p1 should be the function of $\epsilon{_r}$, and it can be called the rate of reduction factor, and p2 must be a constant. The other physical significance of these parameters is yet to be understood.

	\begin{table}[H]
		\center%
		\begin{tabular}{|c|c|c|c|c|}
			\hline 
			d (cm) & p0 & p1 & p2 & p3\tabularnewline
			\hline 
			\hline 
			0.4 & $\begin{array}{c}
			60135\\
			\pm\\
			2189
			\end{array}$ & $\begin{array}{c}
			0.55\\
			\pm\\
			0.03
			\end{array}$ & $\begin{array}{c}
			0.5\\
			\pm\\
			0.02
			\end{array}$ & $\begin{array}{c}
			43511\\
			\pm\\
			129
			\end{array}$\tabularnewline
			\hline 
			0.3 & $\begin{array}{c}
			-68494\\
			\pm\\
			3396
			\end{array}$ & $\begin{array}{c}
			0.74\\
			\pm\\
			0.05
			\end{array}$ & $\begin{array}{c}
			0.43\\
			\pm\\
			0.02
			\end{array}$ & $\begin{array}{c}
			43858\\
			\pm\\
			110
			\end{array}$\tabularnewline
			\hline 
			0.2 & $\begin{array}{c}
			88207\\
			\pm\\
			6462
			\end{array}$ & $\begin{array}{c}
			1.1\\
			\pm\\
			0.07
			\end{array}$ & $\begin{array}{c}
			0.37\\
			\pm\\
			0.02
			\end{array}$ & $\begin{array}{c}
			44230\\
			\pm\\
			81
			\end{array}$\tabularnewline
			\hline 
		\end{tabular}
		
		\caption{\label{table2_fitVAlue}Fit parameters of the plots shown in figure \ref{E_vs_epr}.}
		
	\end{table} 
	
	\subsection{Case 2, variation of d}
\begin{figure}[H]
		\center\subfloat[\label{E_Vs_thickness}]{\includegraphics[scale=0.3]{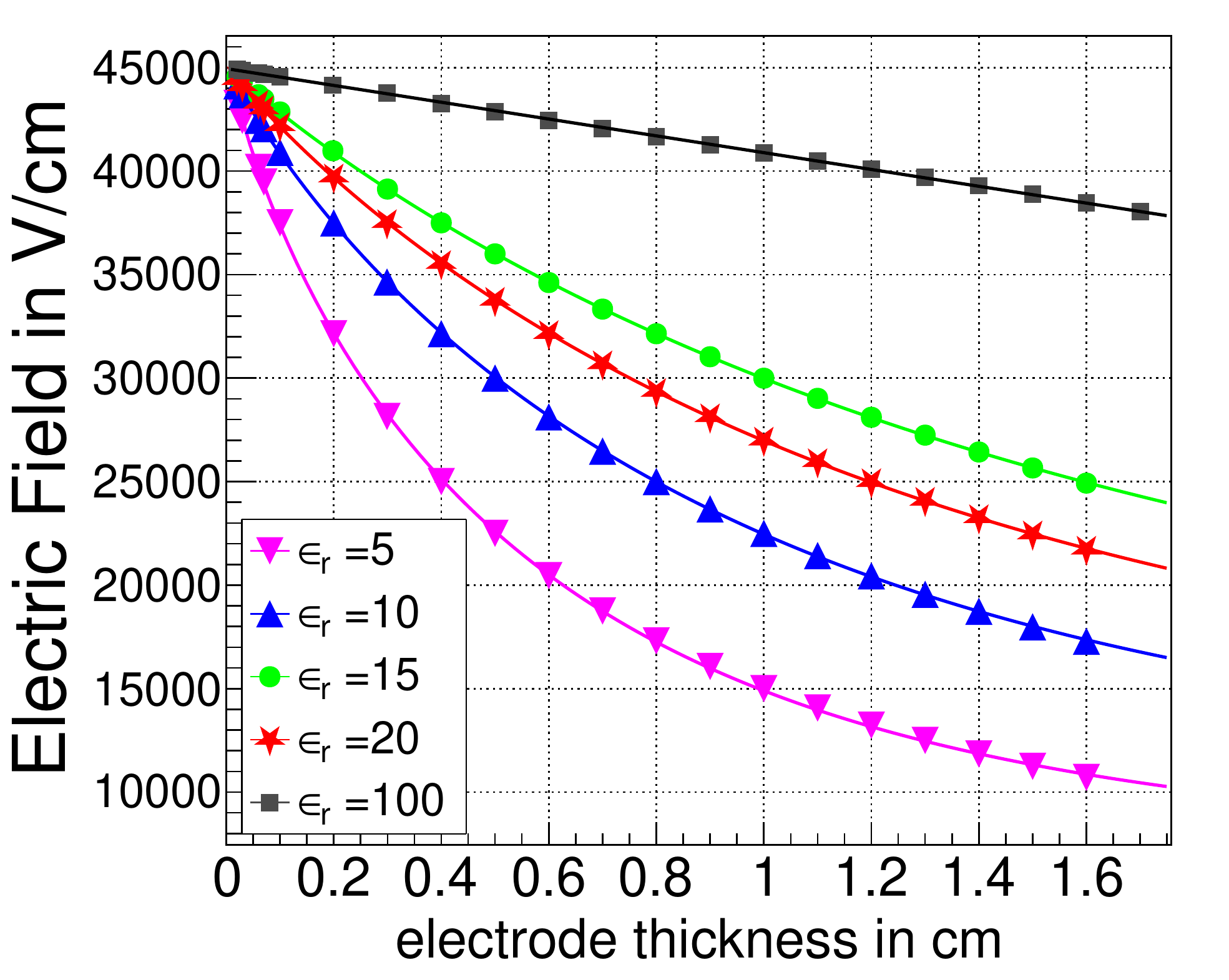}}\subfloat[\label{E_VS_gas_gap}]{\includegraphics[scale=0.3]{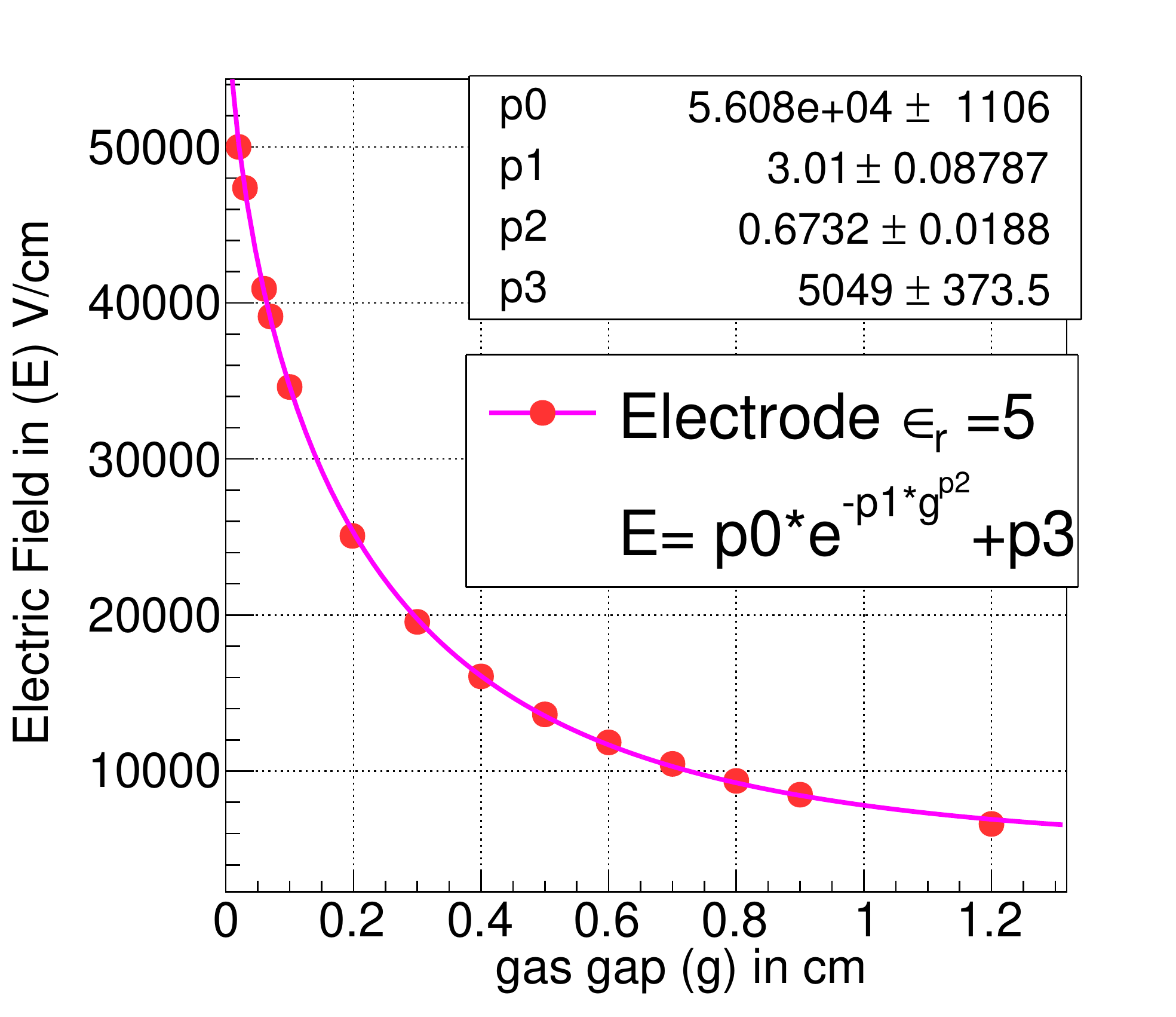}}
		
		\caption{\label{fig:reflection-1-1}(a) Variation of applied electric field with the electrode thickness for different $\epsilon_{r}$ of electrodes. (b) Variation of applied electric field with the gas gap.}
	\end{figure}

	As discussed earlier, electrode thickness is one of the most crucial factors when entering the high particle rate detection region. Therefore, studying the dependence of electric fields on the same is necessary. The variation of the electric field inside the middle of the gas gap with electrode thickness has been shown in figure \ref{E_Vs_thickness}, where the electric field decreases with the increment of the thickness of electrodes. Each curve in figure \ref{E_Vs_thickness} has been shown for a fixed gas gap (g) of 2 mm and several permittivities ($\epsilon_{r}=5,10,15,20,100$), where it may be noted that an increase in $\epsilon_{r}$ leads to flatter curves. This is because the electrodes shift from the dielectric region to the conductive region. Hence, the dependence of the electric field on the electrode thickness for higher $\epsilon_{r}$ is getting small. The maximum percentage of the reduction of the field on the variation of thickness from 0.02 cm to  1.6 cm is shown in the table \ref{epsilon_vs_field_table_3} for several values of $\epsilon_{r}$, where the percentage of change in the field with respect to electrode thickness d is gradually reducing with the increment of $\epsilon_{r}$. Hence, it is expected that for the perfect conductor ($\epsilon_{r}\rightarrow\infty$), the curvature of figure \ref{E_Vs_thickness} will become flat and parallel to the electrode thickness axis. So, the electric field will be independent of the thickness of electrodes.   
	
	\begin{table}[H]
		\center%
		\begin{tabular}{|c|c|}
			\hline 
			$\epsilon_{r}$ & change in field with respect to d (\%)\tabularnewline
			\hline 
			\hline 
			5 & -75.24\%\tabularnewline
			\hline 
			10 & -60.81\%\tabularnewline
			\hline 
			15 & -51.06\%\tabularnewline
			\hline 
			20 & -44.05\%\tabularnewline
			\hline 
			100 & -15.29\%\tabularnewline
			\hline 
		\end{tabular}
		
		\caption{\label{epsilon_vs_field_table_3}Percentage of change in the electric field on variation of thickness from 0.02 cm to 1.6 cm for several permittivity of electrodes ($\epsilon_{r}$). }
		
	\end{table}
	
	\subsection{Case 3, variation of g}
	It is expected that if we fix the electrode thickness and permittivity($\epsilon_{r}$), then the electric field will drop at any point with the increment of the gas gap. In figure \ref{E_VS_gas_gap} the variation of the electric field with the gas gap at the middle of the detector has been shown, where the electrode thickness and permittivity are fixed at d=0.2 cm and $\epsilon_{r}=5$, respectively. The figure \ref{E_VS_gas_gap} has been fitted with the equation: 
	\begin{equation}\label{eqn:2.2}
	f(g)=p0\,exp(-p1\,g^{p2})+p3
	\end{equation}
	 to understand the functional behaviour and convenient interpolation. The fit parameters  p0,p1,p2,p3 has been shown on the same figure \ref{E_VS_gas_gap}. As argued for the parameters of the equation \ref{eqn:2.1} here, the parameters p0 and p3 of the equation \ref{eqn:2.2} carry the dimension of the electric field, and p1 is the function of g and p2 a constant. The other physical significance of these parameters needs further study.

	\section{Calculation of electrode polarisation fields}\label{sec:section3}
	As discussed earlier, the electric field inside an RPC changes while an avalanche is generated. The reason for this change is the growth of space charges inside the RPC. The field due to those space charges can be divided into two sections, (a) field due to the space charges itself (b) field due to the polarisation of electrodes due to those space charges. The electrode polarisation field is calculated using the method of image. In the following subsections, we will first discuss the simplest case, the charged particle inside the two metallic (no dielectric present) grounded plates, and then the discussion will be continued to the case where dielectric layers will be present.
	\subsection{Image charges of a point charge inside two grounded metallic conductor}

	\begin{figure}[H]
		\center\includegraphics[scale=0.5]{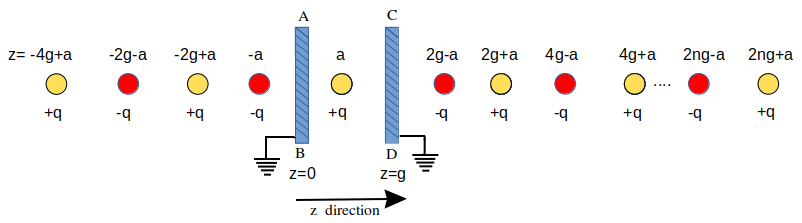}
		
		\caption{\label{fig:reflection}Formation of image charge due to a point charge between two metallic grounded conductors.}
	\end{figure}
\begin{figure}
	
	
	
	\center\subfloat[\label{charge_vs_dist_field_D1}]{\includegraphics[scale=0.31]{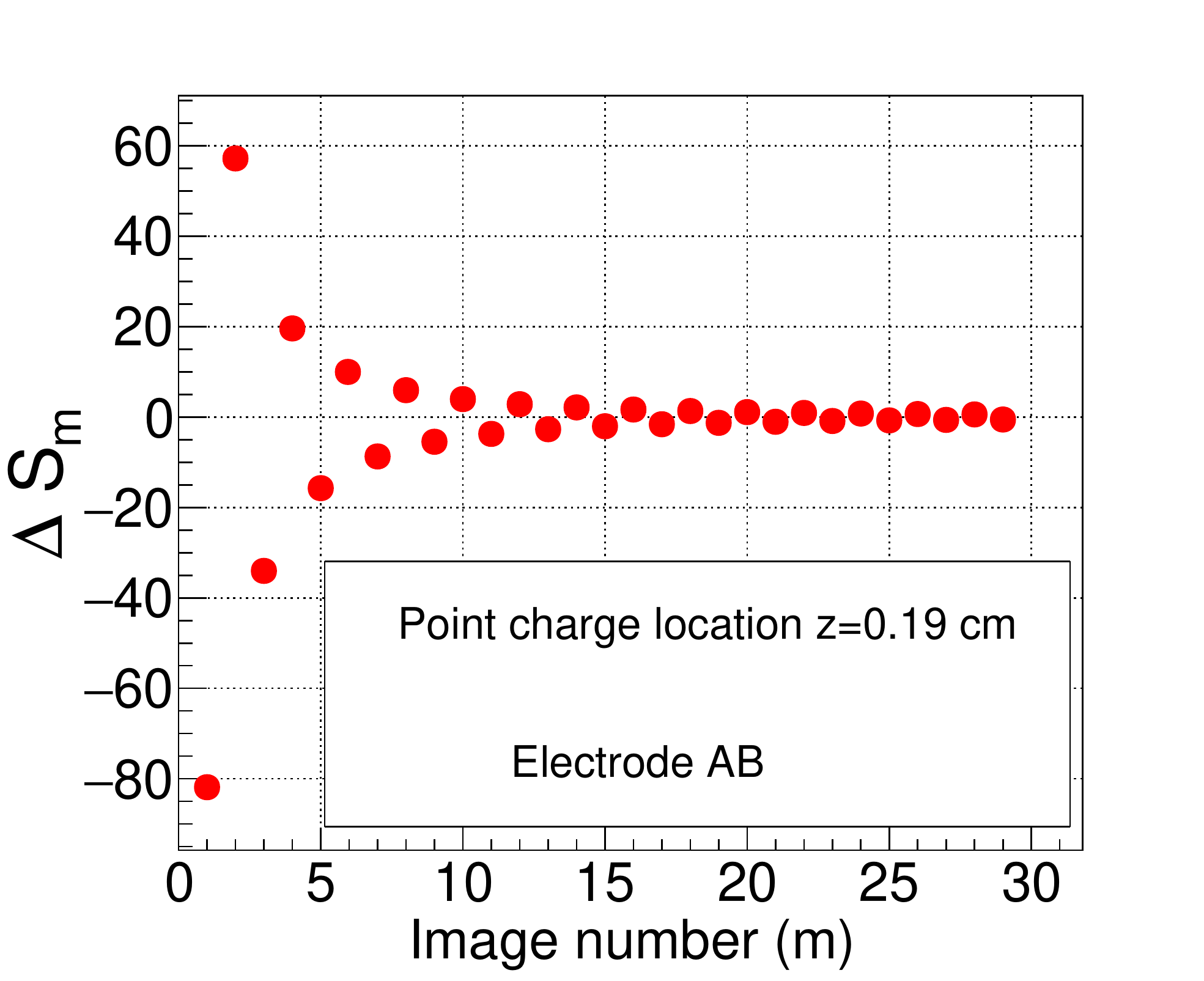}
		
	}\subfloat[\label{charge_vs_dist_field_D2}]{\includegraphics[scale=0.3]{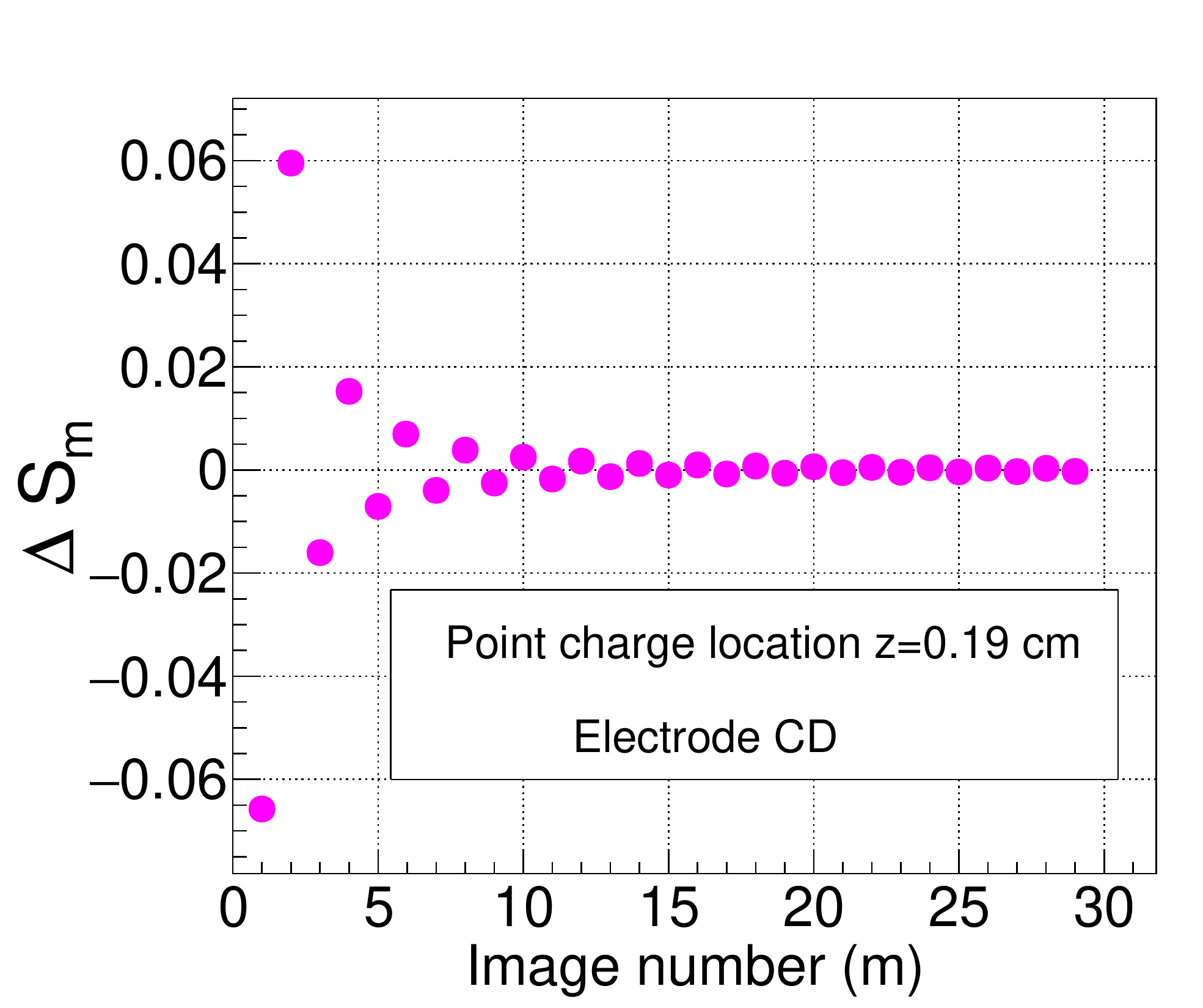}
	}
	
	\caption{(a) Percentage of contribution ($\Delta S_m$) of higher order image charges on the total electric field for electrode AB. (b) Percentage of contribution ($\Delta S_m$) of higher order image charges on the total electric field for electrode CD. }
	
\end{figure}

Let us consider a point charge between two metallic grounded conductors AB and CD, at a distance "a" from the electrode AB (z=0, see figure \ref{fig:reflection}). Now, due to the planar geometry,
there will be an infinite number of reflections of the source
charge on both sides of the
grounded metallic electrodes, as shown in figure \ref{fig:reflection}. Then the electric field at any point between the electrodes is due to the sum of the fields of source charge and infinite image charges. The convergence of infinite series to calculate induced charge density and the total charge for such geometric configuration has been discussed in \cite{new_comb_image_plate}. Since in this work, our focus is on the calculation of electric field, we will find the convergence or termination of infinite series in terms of $E_{n}=\frac{q_{n}}{r_{n}^{2}}$, where $q_{n}=n^{th}$ image charge in electronic charge e unit and $r_{n}=\,\,$distance of $q_{n}$ from nearest electrode and $E_n$ will be different for different electrode. 
In figure \ref{fig:reflection} 
the magnitude and sign of a few image charges corresponding to the electrode AB and CD with their reflection number have been shown, where the magnitudes are the same for all image charges, but signs are alternating. Therefore, the sum of all n number of $E_n$ is $S_{m}=\sum\limits_{n=0}^m E_{n}$. If we multiply $E_{n}$ with coulomb constant and electron charge, then $S_{m}$ will give the total field due to m images on the surfaces of electrodes AB or CD. Hence, the percentage of change in $S_{m-1}$ on the addition of one more image charge is:
\begin{equation}\label{eqn:DeltaSm}
\Delta S_{m}=\frac{S_{m}-S_{m-1}}{S_{m-1}}\times 100
\end{equation}
where, m=1,2,3...., and $\Delta S_{m}$ gives the percentage of contribution of $m^{th}$ order image charge on field. From figure \ref{fig:reflection} we can see that the image charges are alternating between $\pm 1$. Hence, the sign of $\Delta S_{m}$ will also alternate and gradually converges to zero (see figure \ref{charge_vs_dist_field_D1}). For a=0.19, the charge is very close to the electrode CD. Therefore, the term $E_{0}$ (n=0) will be very large for electrode CD with respect to the all other $E_{n}$s (n=1,2,3..). Therefore, in $S_{m}$ the contribution from the higher-order $E_{n}$ (n=1,2,3..) will be small, following that $\Delta S_{m}$ (for m=1,2,3..) will be very small (see figure \ref{charge_vs_dist_field_D2}). On the other hand, the source charge is very far from the electrode AB. Therefore, few higher-order terms of $E_{n}$ (n=1,2,3..) can be comparable to $E_{0}$. 
 This is the reason why $\Delta S_{m}$ is also higher for those few orders (see figure \ref{charge_vs_dist_field_D1}). However, the contribution of images of electrode AB on the entire field is smaller than that of electrode CD. Therefore, while calculating the image field for any electrode, one could apply a cut $\mid\Delta S_{m}\mid$ to terminate the series $S_{m}$ at certain order of reflection. For example, if we cut $\mid\Delta S_{m}\mid\geq 20$, then from figure \ref{charge_vs_dist_field_D1} we can say that we will need to add four higher-order terms and one zeroth-order term to calculate the image field. Using the same cut value we could have only zeroth-order term for electrode CD (figure \ref{charge_vs_dist_field_D2}). Thus, this method opens the opportunity to select several images dynamically throughout the avalanche simulations inside the RPC.	
 \\
  It is noted that the total charge induced on the electrodes due to the infinite images is finite. From figure \ref{fig:reflection}, we can say that the positive charges +q are at z=2ng+a and negative charges -q at z=2ng-a, where n runs from $-\infty$ to $\infty$. If we consider a ring of radius R centered at z-axis, along the line of image charges and width dR then at electrode AB (figure \ref{fig:reflection}) the surface-charge density $\sigma(R)$ can be represented as discussed in \cite{new_comb_image_plate}, which is as follows:
 \begin{eqnarray}\label{eqn:surfaceChargeDensity}
 \sigma(R)&=&\frac{q}{4\pi} \sum_{n=-\infty}^{\infty}[-s(2ng+a,R)+s(2ng-a,R)]
 \end{eqnarray} 
 where
 $s(z,R)=z(R^2+z^2)^{-3/2}$.
The convergence test of infinite series of equation \ref{eqn:surfaceChargeDensity} can be found in \cite{new_comb_image_plate}. 
The total surface charge can be found as follows \cite{new_comb_image_plate}:
\begin{equation}
Q_{AB}=2\pi \int_{0}^{\infty} \sigma (R) RdR=-q\frac{g-a}{g}.
\end{equation} 
Similarly the total charge on the electrode CD can be calculated using the same method. The total surface chharge on CD is $Q_{CD}=-qa/g$. Therefore the total charge on both electrodes AB and CD is $Q=Q_{AB}+Q_{CD}=-q$.
 \subsection{Formation of image charges in two layers of dielectric}
	
	\begin{figure}
		\center\includegraphics[scale=0.5]{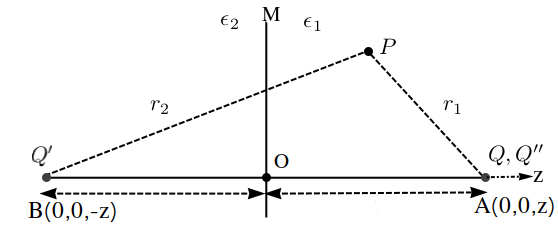}
		
		\caption{\label{fig:polarisation} Formation of image charge for two layers of dielectric case.}
	\end{figure}
	
	Let us consider a point charge $Q$ placed in a dielectric medium of permittivity $\epsilon_{1}$
	at A(0,0,z) from the interface OM of two semi-infinite dielectric medium of permittivity $\epsilon_{1}$ and $\epsilon_{2}$, as shown in figure \ref{fig:polarisation}.
	The field of $Q$ polarises the dielectric and the negative bound charges are
	induced on the surface. The total field at any point P is the sum
	of the field of bound charges and $Q$. Now to calculate potential
	at P we imagine an image charge $Q^{\prime}$ in the dielectric
	medium $\epsilon_{2}$ at position B(0,0,-z) away from the interfacing
	surface of two dielectric medium (figure \ref{fig:polarisation}). Let
	$\phi_{1},\phi_{2}$ denote potential in regions having dielectric permittivities $\epsilon_{1}$and
	$\epsilon_{2}$. Now to satisfy boundary conditions 
	
	\begin{equation}
	\begin{array}{c}
	\phi_{1}|_{z=0}=\phi_{2}|_{z=0}\\
	\epsilon_{1}\frac{\partial\phi_{1}}{\partial z}|_{z=0}=\epsilon_{2}\frac{\partial\phi_{2}}{\partial z}|_{z=0}
	\end{array}\label{eq:boundary_condition}
	\end{equation}
	$Q^{\prime}$should be $Q^{\prime}=-\alpha_{12}Q$
	(where, $\alpha_{12}=\frac{\epsilon_{1}-\epsilon_{2}}{\epsilon_{1}+\epsilon_{2}}$).
	Again to calculate potential at any point $P$ in the medium
	$\epsilon_{2}$ another image charge $Q^{\prime\prime}$ can be
	considered at the point A(0,0,z) in the medium $\epsilon_{1}$. Hence
	to satisfy same boundary condition \ref{eq:boundary_condition} it
	is found that the value of $Q^{\prime\prime}$ should be $Q^{\prime\prime}=\beta_{12}Q$
	(where, $\beta_{12}=\frac{2\epsilon_{2}}{\epsilon_{2}+\epsilon_{1}}$).
	The relation between permittivities can be written as follows to generalise the procedure for multilayered cases:  
	\begin{eqnarray}
	\alpha_{mn}=\frac{\epsilon_{m}-\epsilon_{n}}{\epsilon_{m}+\epsilon_{n}}\label{eqn:alphamn}\\
	\beta_{mn}=\frac{2\epsilon_{n}}{\epsilon_{m}+\epsilon_{n}}
\end{eqnarray}
	where, m=index of source charge and n=index of the reflecting medium. We call $\alpha_{mn}$ as reflection factor and $\beta_{mn}$ as equivalence factor as it defines equivalent charge \cite{Jomaa1983ElectricFD}. $\alpha_{mn}$ uses when we are calculating the field at the same region of the source charge.  $\beta_{mn}$ used to calculate the electric field other than the source charge region. The relation between $\alpha_{mn}$ and $\beta_{mn}$ is below:
	\begin{equation}
	\beta_{mn}=1-\alpha_{mn}.
	\end{equation}
	
	\subsection{Formation of image charges in three layers of dielectric}\label{section5}
	
	\begin{figure}
		\center\includegraphics[scale=0.34]{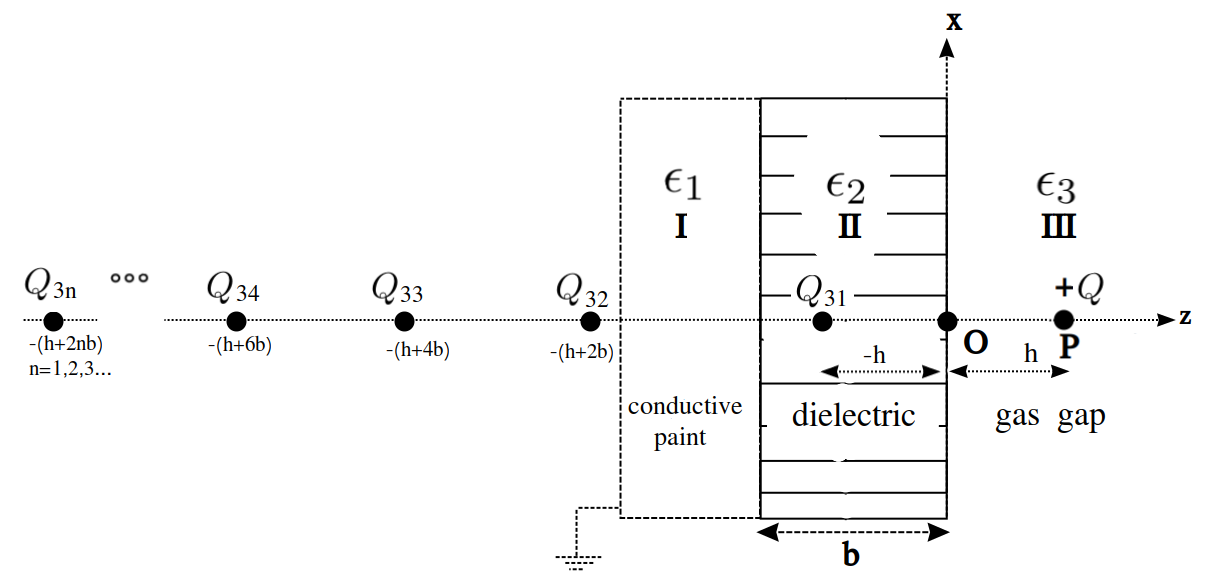}
		
		\caption{\label{fig:threelayer} Formation and positions of image charges due to three layers of dielectric. }
	\end{figure}
	
\begin{figure}
		\center\subfloat[\label{three_dielectric_charge}]{\includegraphics[scale=0.32]{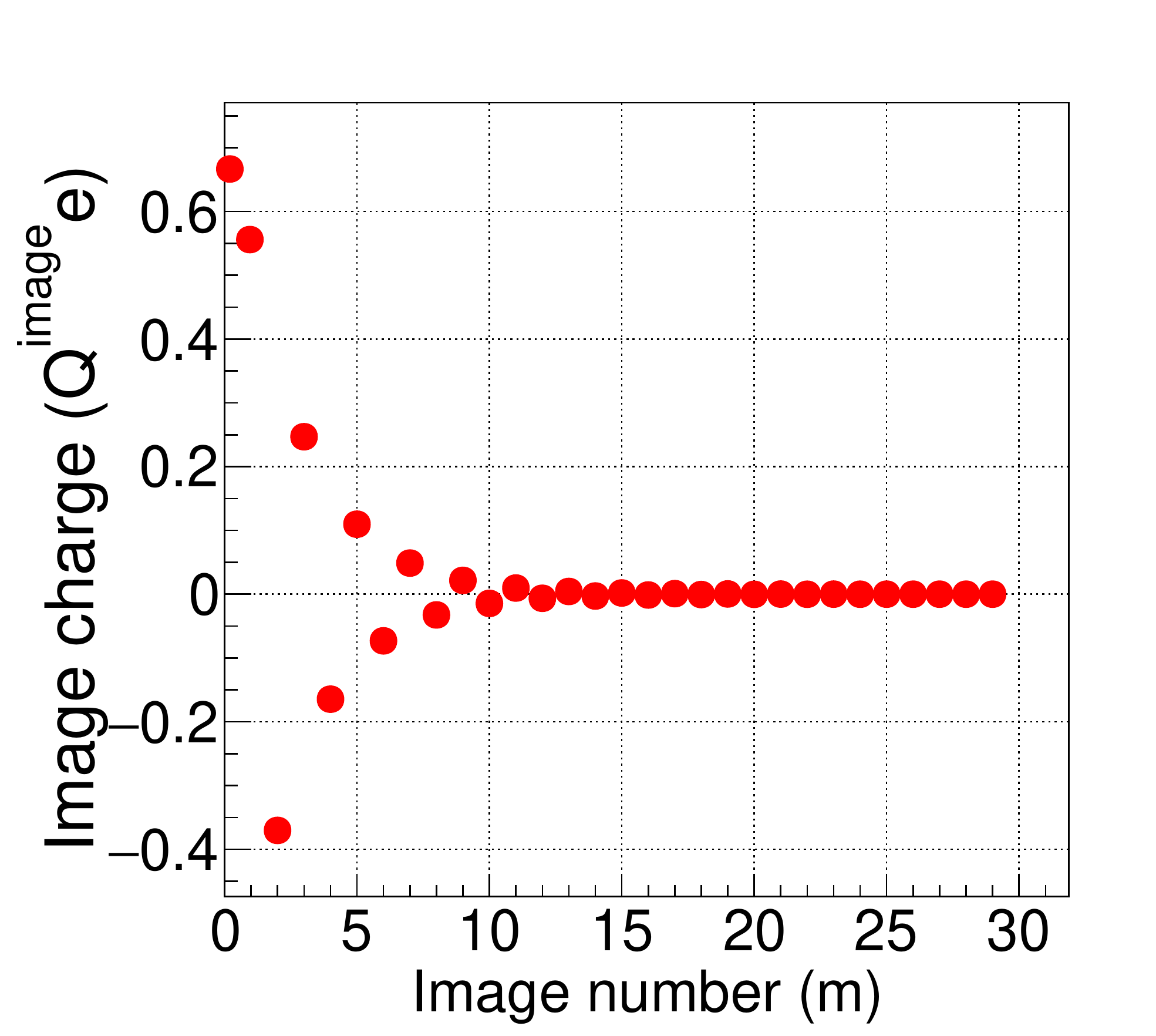}
			
		}\subfloat[\label{three_dielectric_field}]{\includegraphics[scale=0.32]{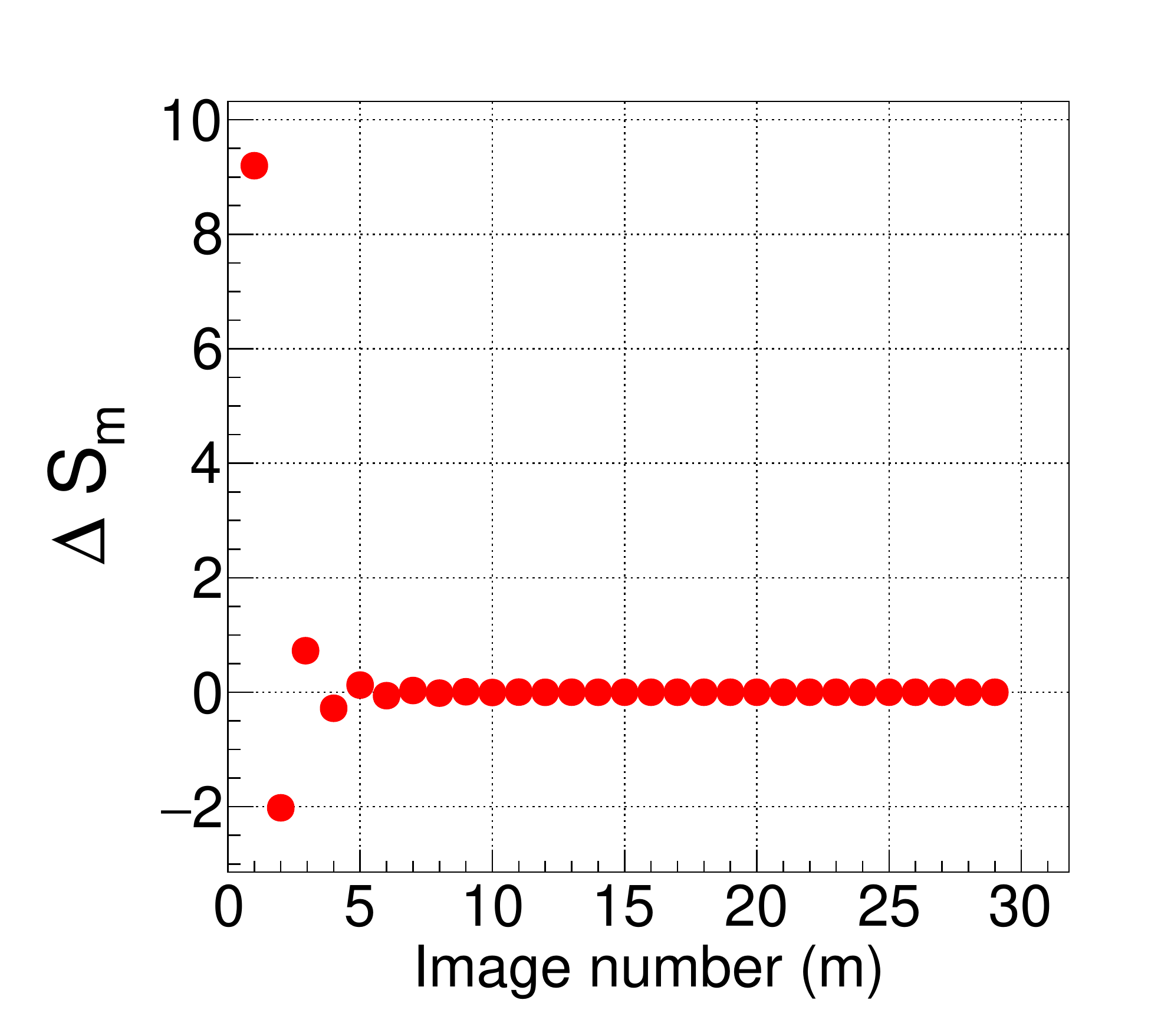}

		}
		
		\caption{\label{fig:charge_intensity-1}(a) Variation of image charges with order of reflection for three layers of dielectric case. (b) Percentage of contribution of higher order image charges on the electric field for three layers of dielectric case.}
	\end{figure}

	Suppose we extend from two layers to three layers. In that case, it can be found that due to two boundary interfaces between medium I and II and between II and III see (figure \ref{fig:threelayer}), an infinite series of image charges is induced\cite{Jomaa1983ElectricFD} due to a source charge +Q placed at a distance h from the interface of dielectric II and III. The series formed by a set of infinite of image charges shown in figure \ref{fig:threelayer} is named as $M3\footnote[1]{The name $M3$ has been chosen since the images are formed in a three-layered medium}$  which can be written as follows:
	 
	 \begin{align}
	M3=\left\lbrace   Q_{31},Q_{32},Q_{33}...Q_{3n}\right\rbrace,
	\end{align}
	where the $n^{th}$ image charge can be written as $Q_{in}$.
	The first index $i$ represents the medium in which we calculate the electric field, and the second index $n$ represents the order of reflection or image charge. Let us consider the thickness of the medium II is b. The positions and charges of the images are shown in the table \ref{tab:threelayer_image}. Here, all distances are measured from the boundary interface between medium III and II. Again to calculate the contribution of higher-order reflections $\Delta S_m$ using equation \ref{eqn:DeltaSm} ( m=2,3..) define $E_{n}=\frac{Q_{in}}{r_n^2}$ and $S_m=\sum_{n=0}^{m}E_n$, where $Q_{in}$ is the $n^{th}$ order image charge and $r_n=-(h+2(n-1)b)$ (for n=1,2,3...) is the distance of $Q_{in}$ from the interface of medium III and II. In figure \ref{three_dielectric_charge} the variation of image charges (for h=0.1mm and b=2mm) with its order of reflection has been shown, where it is clear that the image charges change their sign, and magnitudes gradually reduce with the order. From figure \ref{three_dielectric_field} we can also conclude that the percentage of contribution $\Delta S_m$ of higher-order $m^{th}$ (m=2,3,4..) reflection is also gradually decreasing to zero.  After seven to eight order the value drops from 9.2\% to 0.008\%.    
	
	\begin{table}[H]
		\center%
		\begin{tabular}{|c|c|c|c|}
			\hline 
			Order(n) & Name of charges  & Charge & Position\tabularnewline
			\hline 
			\hline 
			$1^{st}$ & $Q_{31}$ & $\alpha_{32}Q$ & -h\tabularnewline
			\hline 
			$2^{nd}$ & $Q_{32}$ & $(1-\alpha_{32}^{2})\alpha_{21}Q$ & -(h+2b)\tabularnewline
			\hline 
			$3^{rd}$ & $Q_{33}$ & $(1-\alpha_{32}^{2})(-\alpha_{32})\alpha_{21}^{2}Q$ & -(h+4b)\tabularnewline
			\hline 
			&  & $\begin{array}{c}
			.\\
			.\\
			.
			\end{array}$ & \tabularnewline
			\hline 
			$n^{th}$ & $Q_{3n}$ & $(1-\alpha_{32}^{2})(-\alpha_{32})^{n-2}(\alpha_{21})^{n-1}Q$ & -(h+2(n-1)b)\tabularnewline
			\hline 
		\end{tabular}
		
		\caption{\label{tab:threelayer_image}Magnitudes and locations of image charges for the three layer case.}
	\end{table}

	
In our calculation we have considerd that medium-I is a conductor so $\epsilon_{1}$ is considered as infinity. Hence, $\alpha_{21}$ can  be aproximated as below:
\begin{eqnarray}\label{eqn:alpha21}
\alpha_{21}&=&\frac{\epsilon_{2}-\epsilon_{1}}{\epsilon_{2}-\epsilon_{1}}\\\nonumber 
&=&\frac{\frac{\epsilon_{2}}{\epsilon_{1}}-1}{\frac{\epsilon_{2}}{\epsilon_{1}}+1}\\\nonumber
\end{eqnarray}
As $\epsilon_{1}\rightarrow\infty$, so from equation \ref{eqn:alpha21} we can write $\alpha_{21}=-1$.

The sum of all image charges from table	\ref{tab:threelayer_image} can be written as follows:
\begin{eqnarray}
Q_{image}^{Total}&=&Q\left[\alpha_{32}+(1-\alpha_{32}^2)\sum_{n=2}^{\infty}\alpha_{21}^{n-1}(-\alpha_{32})^{n-2} \right]\\\nonumber
&=&Q\left[\alpha_{32}+\frac{(1-\alpha_{32}^2)}{\alpha_{21}\alpha_{32}^{2}}\sum_{n=2}^{\infty}\alpha_{21}^{n}(-\alpha_{32})^{n}\right]\\\nonumber
&=&Q\left[\alpha_{32}+\frac{(1-\alpha_{32}^2)}{\alpha_{21}\alpha_{32}^{2}}\Bigg\{\sum_{n=0}^{\infty}[\alpha_{21}^{n}(-\alpha_{32})^{n}]-(1-\alpha_{21}\alpha_{32})\Bigg\}\right]\\\nonumber
&=&Q\left[\alpha_{32}-\frac{(1-\alpha_{32}^2)}{\alpha_{32}^{2}}\Bigg\{\sum_{n=0}^{\infty}[(-1)^{2n}\alpha_{32}^{n}]-(1+\alpha_{32})\Bigg\}\right],(As\; \alpha_{21}=-1)\\ \nonumber
&=&Q\left[\alpha_{32}-\frac{(1-\alpha_{32}^2)}{\alpha_{32}^{2}}\Bigg\{\frac{1}{1-\alpha_{32}}-(1+\alpha_{32})\Bigg\}\right],(As\; \sum_{n=0}^{\infty}\alpha_{32}^{n}=\frac{1}{1-\alpha_{32}},\,and \mid\alpha_{32}\mid<1)\\ \nonumber
&=&-Q
\end{eqnarray}

	\subsection{Formation of image charges in RPC} \label{sec:sec_6}
\begin{figure}
		\center\includegraphics[scale=0.27]{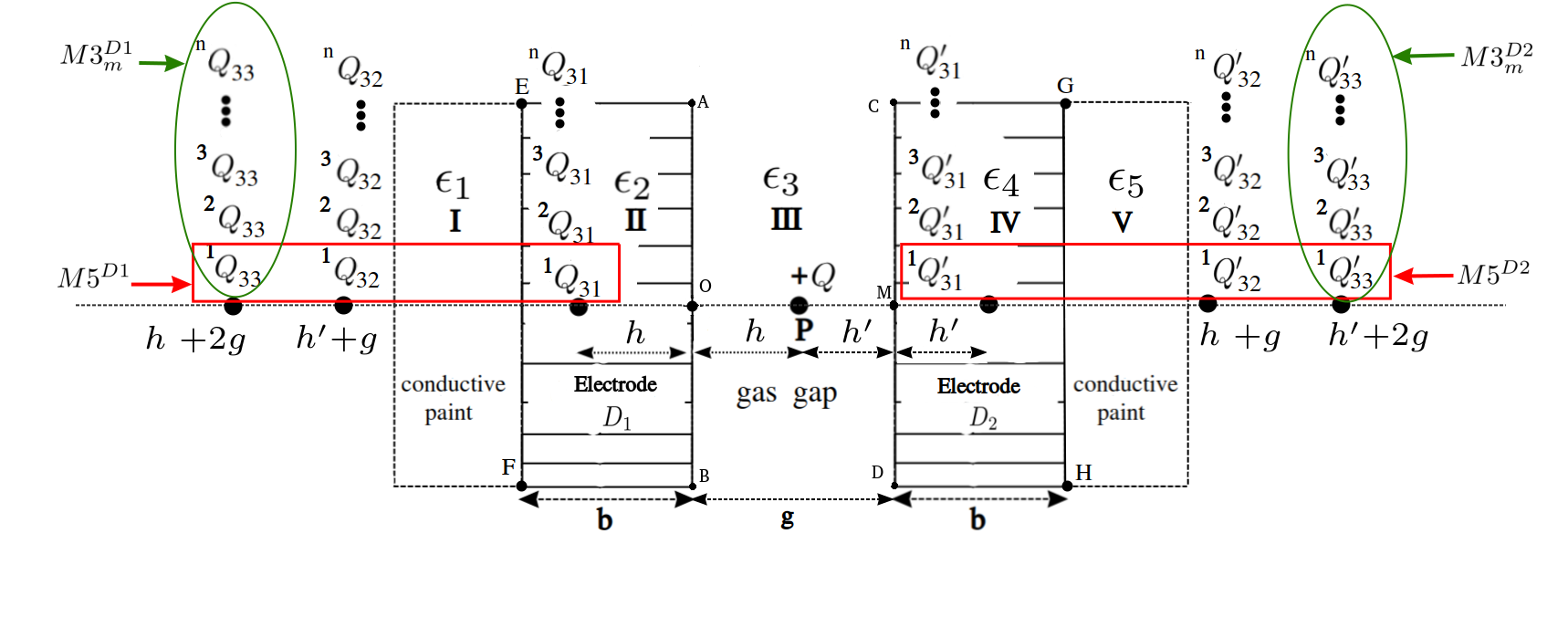}
		
		\caption{\label{fig:RPC_image_charge_formation}Formation of image charges in an RPC.}
	\end{figure}

	It is known that RPC contains two dielectric electrodes along with a layer of conductive graphite paint (see figure \ref{fig:RPC_image_charge_formation}). For now, we are considering graphite paint as a perfect conductor. Due to the dielectric presence, the image's formation will be different from the metal electrode case, as described below.

\subsubsection{Formation of $M3$ series for electrodes D1 and D2}\label{subsec:6.1}	
Let us consider a charge Q located inside the gas gap (medium III) at a distance of $h$ from the left electrode D1 and $h^{\prime}$ from the inner surface of the right electrode D2 (see figure \ref{fig:RPC_image_charge_formation}). To calculate the position of image charges, we divide the whole system of figure \ref{fig:RPC_image_charge_formation} such that there are three layers of the medium on the left side (I, II, III) and three on the right (III, IV, V). In section \ref{section5} for the three-layer case (see figure \ref{fig:threelayer}), we have seen a series $M3$ of infinite image charges formed due to a single point charge. Similarly, we will also get two similar kinds of $M3$ series for three layers of medium-I, II, III, and medium-III, IV, V (see figure \ref{fig:RPC_image_charge_formation}). Since the left and right $M3$ series of figure \ref{fig:RPC_image_charge_formation} correspond to electrodes D1 and D2, respectively; hence the left and right $M3$ series can be written as follows:
 
 \begin{align}
M3^{D1}=\left\lbrace   ~^1\!Q_{31},~^2\!Q_{31},~^3\!Q_{31}...~^n\!Q_{31}\right\rbrace \\ 
 M3^{D2}=\left\lbrace ~^1\!Q_{31}^\prime,~^2\!Q_{31}^\prime,~^3\!Q_{31}^\prime...~^n\!Q_{31}^\prime\right\rbrace,
 \end{align}
  Where the index at superscript ($n$)  represents the order of reflection and the prime symbol ($~^1\!Q_{31}^\prime$) on image charges of D2  has been used to separate from image charges of D1. The significance of all indices at the subscript will be discussed in the next section. As discussed in section \ref{section5} the value of $\Delta S_{m}$ can be used to decide the significant number of terms of series $M3^{D1, D2}$.
\subsubsection{Formation of image charges in five layers of dielectric}\label{subsec:6.2}

If we consider 1st order reflection $ ~^1\!Q_{31} $ as a source charge for electrode D2, it will generate another $M3^{D2}$ series of second-order at D2, which is $\left\lbrace ~^1\!Q_{32}^\prime,~^2\!Q_{32}^\prime,~^3\!Q_{32}^\prime...~^n\!Q_{32}^\prime\right\rbrace$.  Similarly, if we take $~^1\!Q_{31}^\prime$  as a source charge for D1, it will generate another  $M3^{D1}$ series at electrode D1, which is $\left\lbrace   ~^1\!Q_{32},~^2\!Q_{32},~^3\!Q_{32}...~^n\!Q_{32}\right\rbrace $. This process will continue infinitely and generate two series of infinite images for electrodes D1 and D2. As the formation of these series corresponds to five layers of the dielectric medium, we named this $M5$. The mathematical representation of $M5$ for D1 and D2 is given below,
 
 \begin{align}
 M5^{D1}=\left\lbrace \left\lbrace ~^1\!Q_{31}..~^n\!Q_{31} \right\rbrace , \left\lbrace ~^1\!Q_{32}..~^n\!Q_{32} \right\rbrace,\left\lbrace ~^1\!Q_{33}..~^n\!Q_{33} \right\rbrace ...\left\lbrace ~^1\!Q_{3m}..~^n\!Q_{3m} \right\rbrace  \right\rbrace \\
  M5^{D2}=\left\lbrace \left\lbrace ~^1\!Q_{31}^\prime..~^n\!Q_{31}^\prime \right\rbrace  , \left\lbrace  ~^1\!Q_{32}^\prime..~^n\!Q_{32}^\prime \right\rbrace , \left\lbrace  ~^1\!Q_{33}^\prime..~^n\!Q_{33}^\prime \right\rbrace ...\left\lbrace  ~^1\!Q_{3m}^\prime..~^n\!Q_{3m}^\prime \right\rbrace  \right\rbrace,
  \end{align} 
   where $M5^{D1}$, $M5^{D2}$ stands for the electrode D1, D2 respectively,  m is the order of the  $M5^{D1,D2}$ series and the index 3 of the charges ($~^n\!Q_{3m},~^n\!Q_{3m}^\prime$) is corresponding to the medium at which we are calculating the electric field. 
   Alternatively, in a composite form for D1 and D2, we can write,
  
   \begin{equation}
   M5^{D1,D2}=\left\lbrace M3^{D1,D2}_{1},M3^{D1,D2}_{2},M3^{D1,D2}_{3}...M3^{D1,D2}_{m} \right\rbrace
   \end{equation}
   where, $M3_m^{D1,D2}=$ $\left\lbrace ~^1\!Q_{3m}..~^n\!Q_{3m} \right\rbrace$ or $\left\lbrace  ~^1\!Q_{3m}^\prime..~^n\!Q_{3m}^\prime \right\rbrace$, and it represents $m^{th}$ order of  $M3^{D1,D2}$ series for D1 or D2. Since the higher-order reflections $~^2\!Q_{3m},~^3\!Q_{3m}...~^n\!Q_{3m}$ or $~^2\!Q_{3m}^\prime,~^3\!Q_{3m}^\prime...~^n\!Q_{3m}^\prime$ etc. are at considerably far from the electrode D1 and D2, so their image charge will be generated even further from both the electrodes. Hence the effect of images of them in the field calculations is negligible
and we only consider the images of $~^1\!Q_{3m}$ and $~^1\!Q_{3m}^\prime$. Therefore now the $M5^{D1,D2}$ becomes:
 
  \begin{align}
 M5^{D1}=\left\lbrace  ~^1\!Q_{31}  ,  ~^1\!Q_{32} , ~^1\!Q_{33} ... ~^1\!Q_{3m} \right\rbrace \\
 M5^{D2}=\left\lbrace ~^1\!Q_{31}^\prime   , ~^1\!Q_{32}^\prime  ,  ~^1\!Q_{33}^\prime  ...  ~^1\!Q_{3m}^\prime \right\rbrace.
 \end{align} 

Indeed, one can include some of the higher-order reflections in the field calculation according to the need for precision but not their images.

The magnitude and location of all charges in $M5^{D1, D2}$ series have been shown in table \ref{tab:genaration of image RPC}. The arrow's tail in table \ref{tab:genaration of image RPC} denotes the source charge, and the head indicates the image charge corresponding to the source. $\alpha_{32}$ and $\alpha_{34}$ is the reflection factor of D1 and D2, respectively.
The series shown in table \ref{tab:genaration of image RPC}  can be divided into even and odd series based on their order of reflection for ease of calculations, which is represented in table \ref{tab:even_odd_series_table6}. It is noted that the $h$ and $h^{\prime}$ are just the magnitude of the distance from the inner surface of electrodes D1 and D2. Hence, to find the real and image charge locations about a fixed origin, one will need to use suitable coordinate transformations. The algorithm of the generation of image charge for D1 electrode has been discussed in the appendix \ref{appendix}. The same method can be applied to generate the image for the D2 electrode.		
	\begin{table}
		\center\includegraphics[scale=0.35]{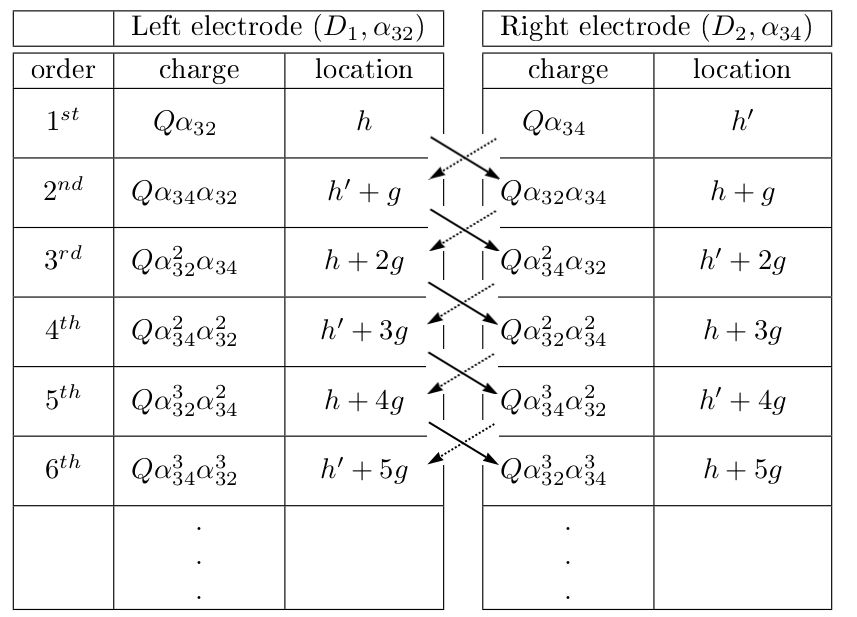}
		
		\caption{\label{tab:genaration of image RPC}Locations of image charges of $M5^{D1,D2}$, where the direction of arrow denotes source to image charge.}
	\end{table}
	\begin{table}
		\center%
		\begin{tabular}{|>{\centering}p{2cm}|>{\centering}p{2cm}|>{\centering}p{2cm}|}
			\hline 
			\multicolumn{3}{|c|}{Odd Series $(Qp=Q \alpha_{32})$}\tabularnewline
			\hline 
			\hline 
			Name & Charge & Location\tabularnewline
			\hline 
			$Q_{11}$$\begin{array}{c}
			\\
			\\
			\end{array}$ & $Qp$ & $h$\tabularnewline
			\hline 
			$Q_{13}$$\begin{array}{c}
			\\
			\\
			\end{array}$ & $Qp\,\alpha_{32}\alpha_{34}$ & $h+2g$\tabularnewline
			\hline 
			$Q_{15}$$\begin{array}{c}
			\\
			\\
			\end{array}$ & $Qp\,\alpha_{32}^{2}\alpha_{34}^{2}$ & $h+4g$\tabularnewline
			\hline 
			& $\begin{array}{c}
			.\\
			.
			\end{array}$ & \tabularnewline
			\hline 
			$Q_{1(2n_{1}+1)}$$\begin{array}{c}
			\\
			\\
			\end{array}$ & $Qp\,\alpha_{32}^{n_{1}}\alpha_{34}^{n_{1}}$ & $h+2n_{1}g$\tabularnewline
			\hline 
		\end{tabular}~~%
		\begin{tabular}{|>{\centering}p{2cm}|>{\centering}p{2.5cm}|>{\centering}p{2.6cm}|}
			\hline 
			\multicolumn{3}{|c|}{Even Series $(Qp=Q \alpha_{32})$}\tabularnewline
			\hline 
			\hline 
			Name & Charge & Location\tabularnewline
			\hline 
			$Q_{12}$$\begin{array}{c}
			\\
			\\
			\end{array}$ & $Qp\,\alpha_{34}$ & $2g-h$\tabularnewline
			\hline 
			$Q_{14}$$\begin{array}{c}
			\\
			\\
			\end{array}$ & $Qp\,\alpha_{34}^{2}\alpha_{32}$ & $4g-h$\tabularnewline
			\hline 
			$Q_{16}$$\begin{array}{c}
			\\
			\\
			\end{array}$ & $Qp\,\alpha_{34}^{3}\alpha_{32}^{2}$ & $6g-h$\tabularnewline
			\hline 
			& $\begin{array}{c}
			.\\
			.
			\end{array}$ & \tabularnewline
			\hline 
			$Q_{1(2n_{2})}$$\begin{array}{c}
			\\
			\\
			\end{array}$ & $Qp\,\alpha_{34}^{n_{2}+1}\alpha_{32}^{n_{2}}$ & $2(n_{2}+1)g-h$\tabularnewline
			\hline 
		\end{tabular}
		
		\caption{\label{tab:even_odd_series_table6}Division of table \ref{tab:genaration of image RPC} into odd and even series.}
	\end{table}
		\subsubsection{ Image charge selection criteria for an RPC }\label{subsec:6.3}
Since the images are symmetric for electrode D1 and D2, we will show results for D1 only. Unlike metal electrodes, in the case of an RPC, $m^{th}$ order of reflection is a sum of n number of image charges of  $M3^{D1}_m$ series(see figure \ref{fig:RPC_image_charge_formation}). Therefore, we can define $ E_{j}$ as, $E_{j}=\sum\limits_{i=1}^{n}\frac{^i\!Q_{3j}}{r_{i}^{2}}$, where $r_{i}$= position of the image charge $^i\!Q_{3j}$ measured from the corresponding electrode and n is the order of  $M3^{D1}_m$ series. Since we have neglected the higher order terms of $M3_m^{D1}$; hence in this case the value of n is 1. If m is the order of the  $M5^{D1}$, till which effects are considered, then we can write ${S}_{m}=\sum\limits_{j=1}^{m}{E}_{j}$. Therefore, the value of $\Delta S_{m}$ can be calculated using equation \ref{eqn:DeltaSm}, where m = 2,3,4....
\begin{figure}[h]
		\center\subfloat[\label{RPCcharge_vs_dist_D1}]{\includegraphics[scale=0.3]{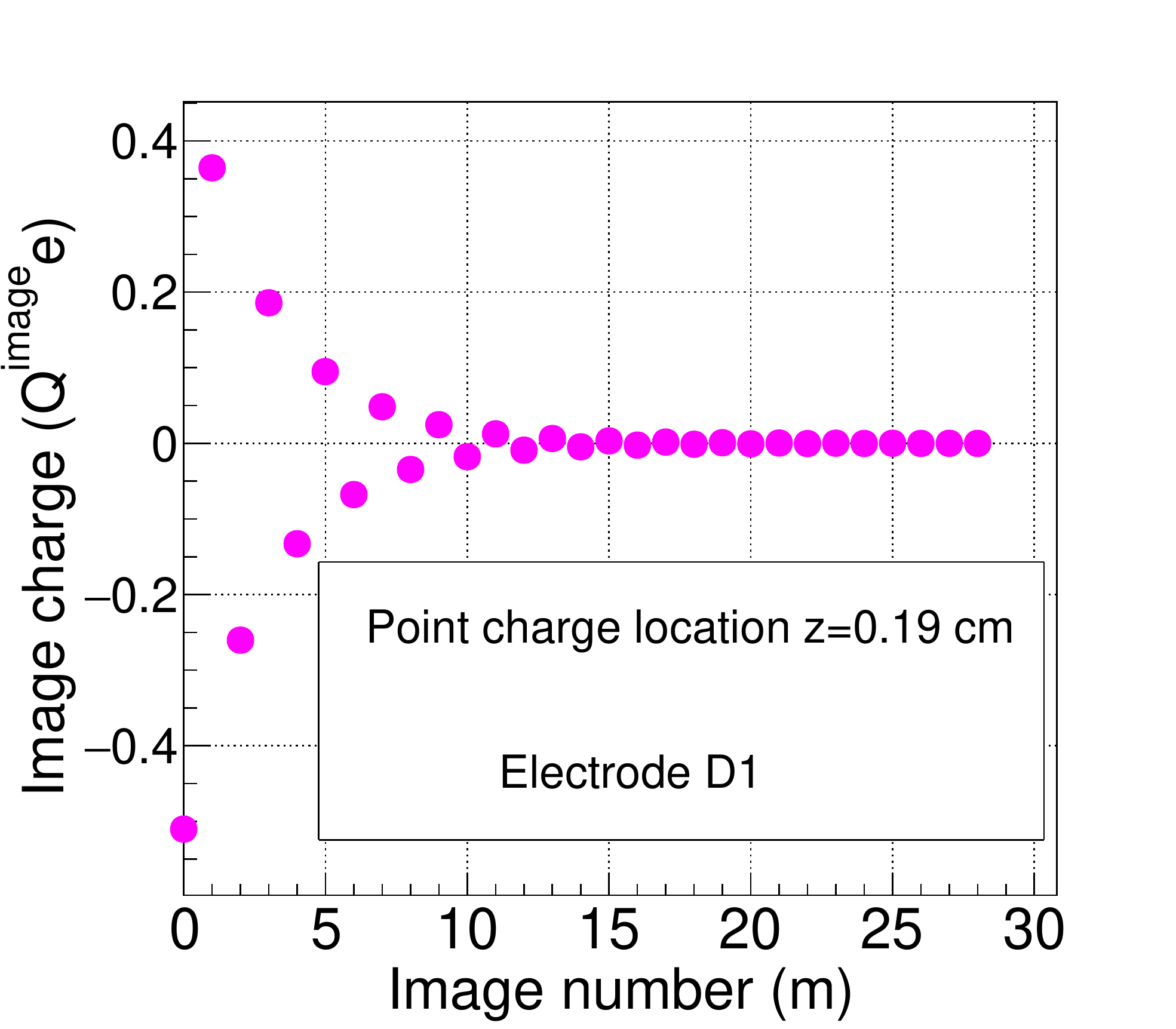}
			
		}\subfloat[\label{RPCcharge_vs_dist_D2}]{\includegraphics[scale=0.32]{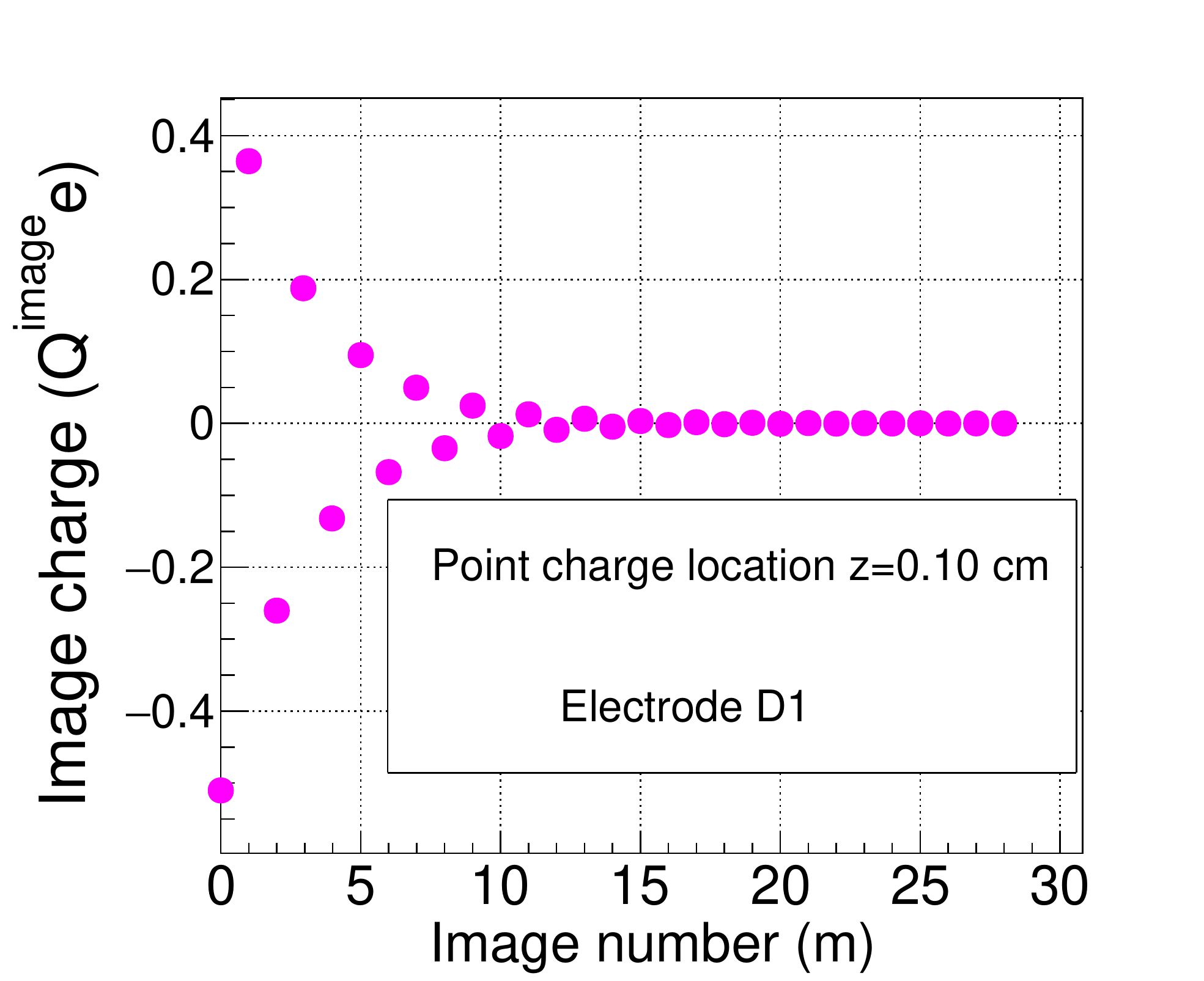}
			
		}
		
		\center\subfloat[\label{RPCcharge_vs_dist_field_D1}]{\includegraphics[scale=0.31]{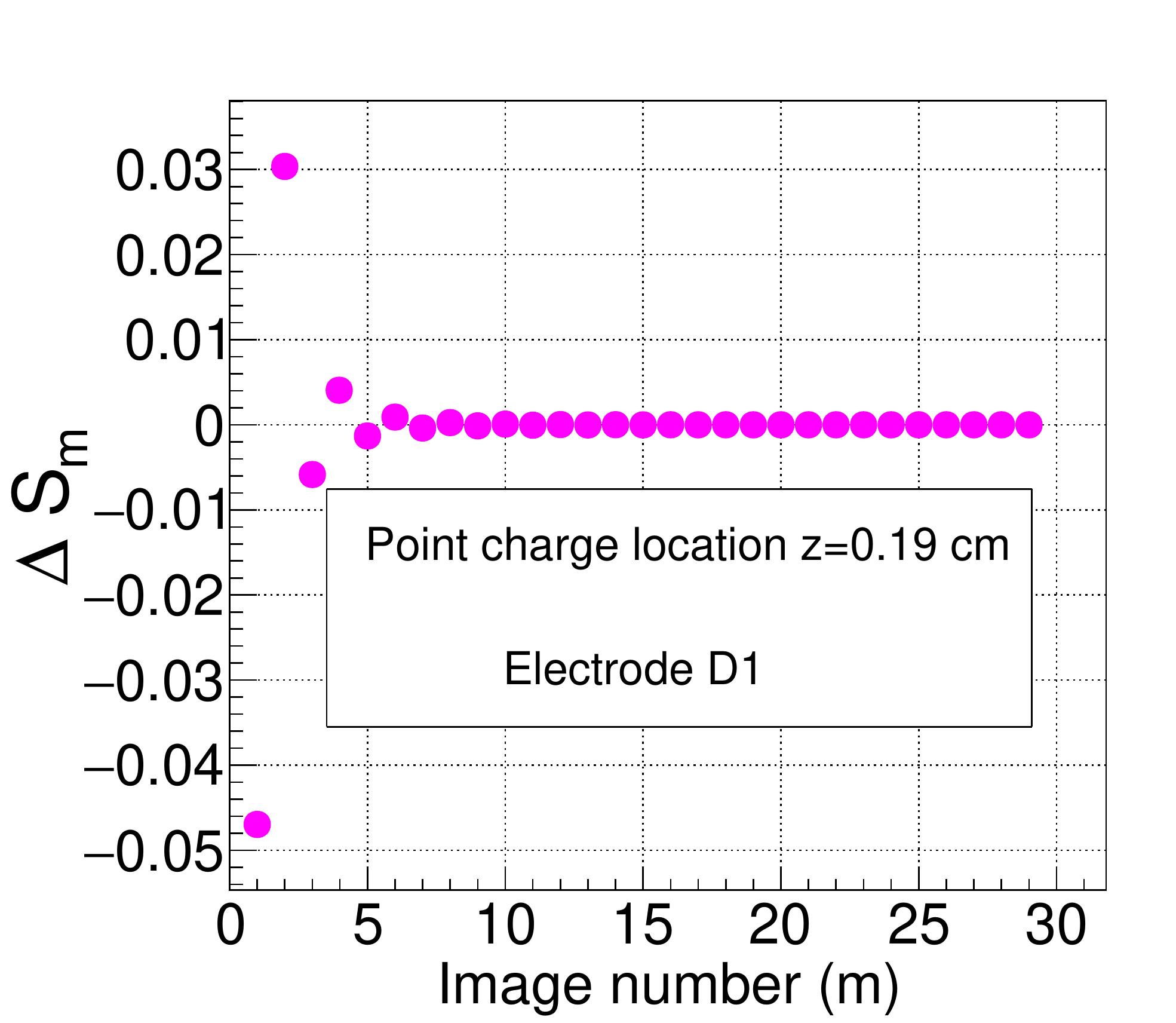}
			
		}\subfloat[\label{RPCcharge_vs_dist_field_D2}]{\includegraphics[scale=0.32]{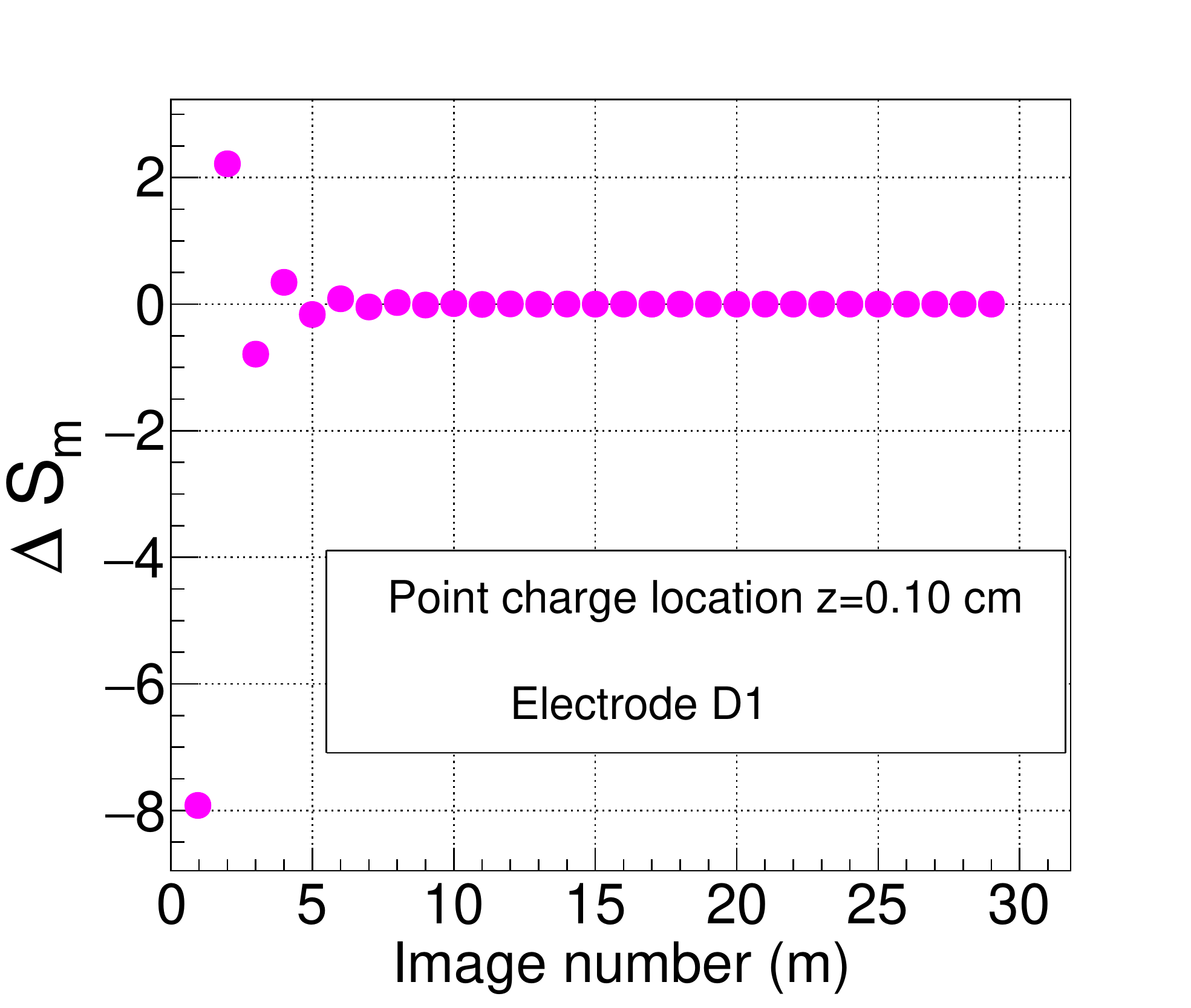}
		}
		
		\caption{(i) Variation of image charge with the order of reflection for electrode D1 of an RPC,when the point charge location at (a) z=0.019 cm and (b) z=0.1 cm. (ii) Percentage of contribution of
		higher order image charges on the electric field, when the point charge is located at (c) z=0.019 cm and (d) z=0.1 cm. }
		
	\end{figure}
Let us now consider electrode thickness b = 2 mm and gas-gap g = 2 mm. The magnitude and sign of the image charges of $M5^{D1}$ series due to a point source charge located at two different positions a) near to the electrode D1 (0,0,0.19) and b) middle of the gas-gap (0,0,0.1), has been shown in figures \ref{RPCcharge_vs_dist_D1} and \ref{RPCcharge_vs_dist_D2}, where the origin is at the inner surface of electrode D2. The sign of image charges alternates between plus and minus, and their magnitude gradually decreases with order increment. The contribution in-field evaluation of few terms from the  $M5^{D1}$ series for two different locations of the source charge has been shown in figures \ref{RPCcharge_vs_dist_field_D1} and \ref{RPCcharge_vs_dist_field_D2}. 
	It is found that the maximum contribution of the higher-order term of  $M5^{D1}$ series is nearly 8$\%$ and continuously converges to zero as order increases.\\
	The sum of image charges of $M5^{D1}$ series can be calculated from the table \ref{tab:even_odd_series_table6} as follows:
	
	\begin{eqnarray}\label{eqn:imageChargeD1}
	Q_{Total}^{D1}&=&(Q_{11}+Q_{13}+Q_{15}...) +(Q_{12}+Q_{14}+Q_{16}...)\\\nonumber
	&=&Q\,\alpha_{32}\sum_{n_1=0}^{\infty}[(\alpha_{32}\,\alpha_{34})^{n_{1}}]+Q \alpha_{32}\,\alpha_{34}\sum_{n_2=0}^{\infty}[(\alpha_{32}\,\alpha_{34})^{n_2}]\\\nonumber
	&=&\frac{Q\,\alpha_{32}}{1-\alpha_{32}\,\alpha_{34}}+\frac{Q\,\alpha_{32}\,\alpha_{34}}{1-\alpha_{32}\,\alpha_{34}},(As,\; \sum_{n=0}^{\infty}(\alpha_{32}\,\alpha_{34})^{n}=\frac{1}{1-\alpha_{34}\,\alpha_{32}}, and \, \mid\alpha_{32}\,\alpha_{34}\mid<1)\\\nonumber
	&=&Q\alpha_{32}\frac{1+\alpha_{34}}{1-\alpha_{32}\,\alpha_{34}}
	\end{eqnarray}  
	Let us consider that the permittivity of D1 and D2 is the same. Hence, $\alpha_{32}=\alpha_{34}$  and so from the equation \ref{eqn:imageChargeD1} we can write:
	\begin{eqnarray}\label{eqn:Qtot_D1}
	Q_{Total}^{D1}=\frac{Q\,\alpha_{32}}{1-\alpha_{32}}.
	\end{eqnarray}
	It is noted that the medium III is gas; hence the permittivity of it is equivalent to free space permittivity ($\epsilon_{0}$) and so $\epsilon_{3}=\epsilon_{0}$. The reflection factor $\alpha_{32}$ can be written as using equation \ref{eqn:alphamn}:
	\begin{eqnarray}\label{eqn:alpha32}
	\alpha_{32}&=& \frac{\epsilon_{3}-\epsilon_{2}}{\epsilon_{3}+\epsilon_{2}}\\\nonumber
	&=&\frac{\epsilon_{0}-\epsilon_{2}}{\epsilon_{0}+\epsilon_{2}}.
	\end{eqnarray}
	Now as $\epsilon_{0}<\epsilon_{2}$; hence $\alpha_{32}<0$. Therefore, using equation \ref{eqn:Qtot_D1} we can say that the sign of $Q_{Total}^{D1}$ is negative and $\mid Q_{Total}^{D1}\mid<\mid Q\mid$. Similarly we can find the total image charges for $D2$ electrodes.

	\section{Calculation of the space charge field  and image field of avalanche inside an RPC} \label{sec:Section4_avalancheCharge}
		\begin{figure}
		\center\subfloat[\label{avalanche_electron_dist}]{\includegraphics[scale=0.4]{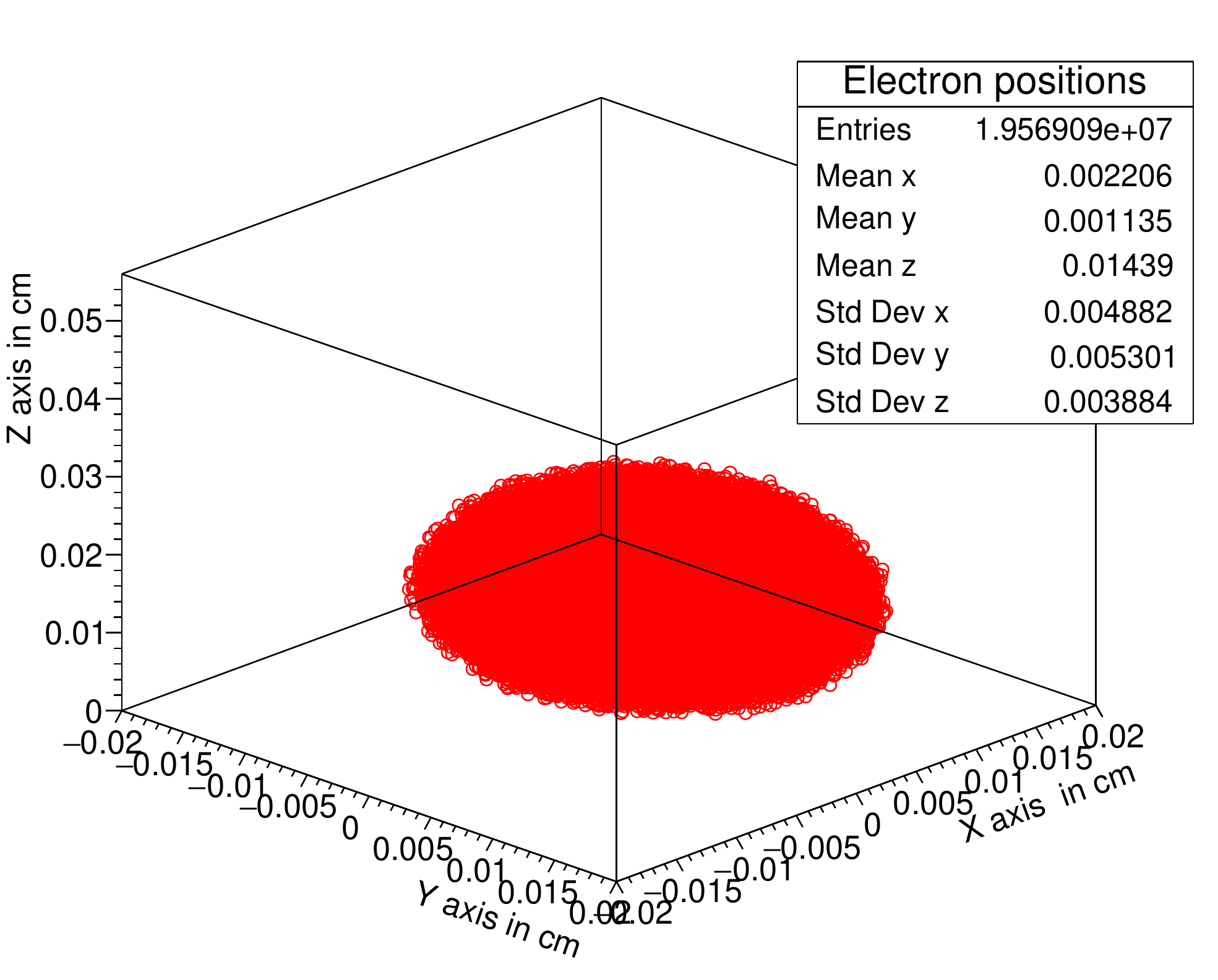}
			
		}\subfloat[\label{avalanche_ion_dist}]{\includegraphics[scale=0.4]{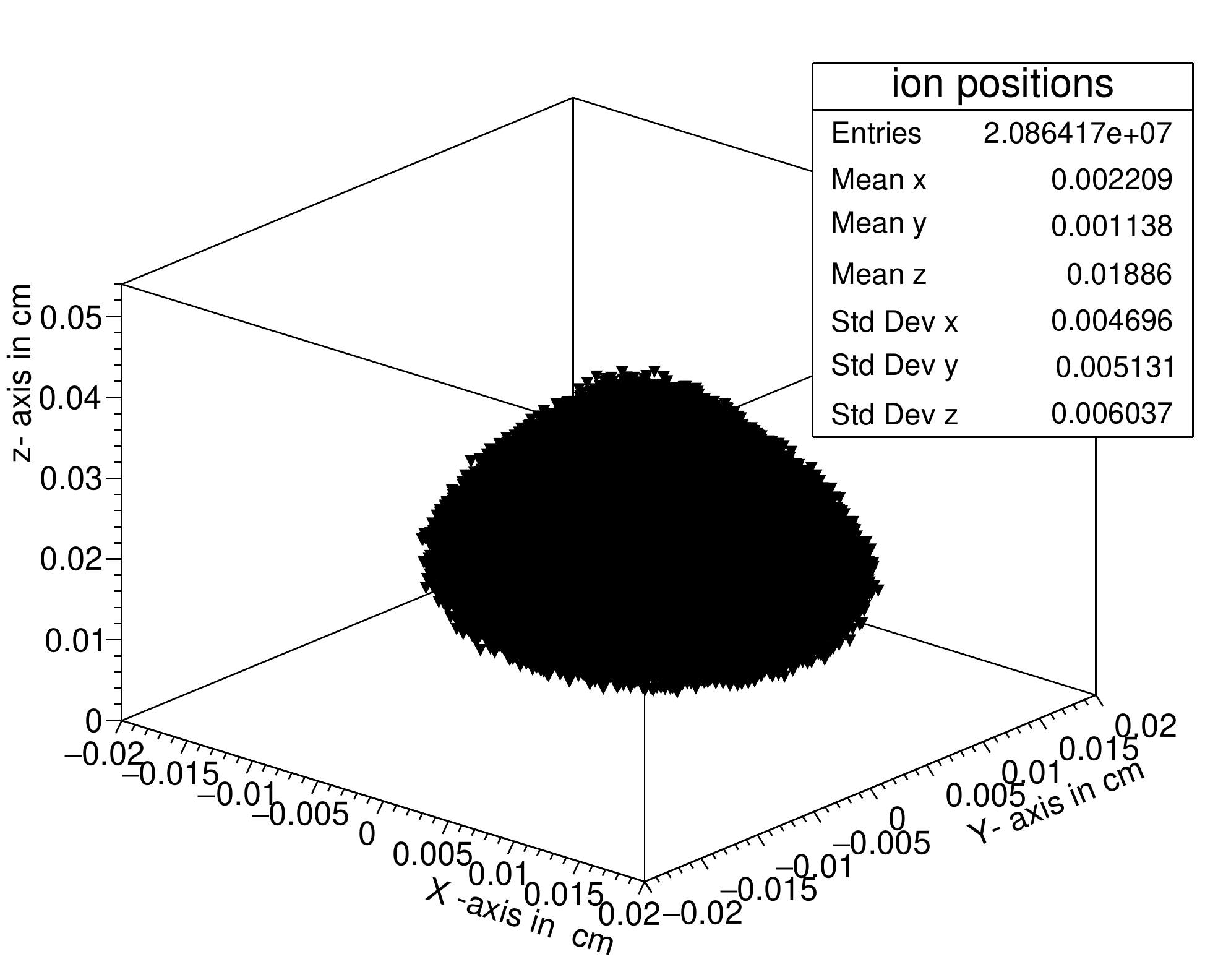}
			
		}

		\caption{(a) Simulated avalanche electron distribution at certain instant of time. (b) Simulated avalanche ion distribution at a certain instant of time. }
	\end{figure}
	It is known that the generation of space charge inside the RPC is high when the incoming particle rate is high. This condition also can be achieved with a single primary electron if the applied field is sufficiently high for a particular gas mixture. An avalanche charge distribution at an instant of time has been simulated from a single primary electron inside an RPC using Garfield++, where the geometry and applied voltage discussed in section \ref{sec:section2} is used, and the initial position of the electron has been chosen at the centre of the RPC. The thickness of the electrode and gas gap is fixed at 2 mm. The drift velocity has been calculated using MAGBOLTZ \cite{BIAGI1989716} to introduce drift motion in the simulation. For the thermal diffusive motion, we have considered that the diffusion is described as Gaussian distribution. It is known that under the electric field, the gaussian diffusion becomes anisotropic and so the distribution becomes \cite{LippmanThesis}:
	\begin{eqnarray}
	\phi_L(z,l)&=&\frac{1}{\sqrt{2\pi l}D_L}exp\Big(-\frac{(z-z_0)^2)}{2D_L^2 l}\Big)\\
	\phi_T(r,l)&=&\frac{1}{D^2_T l}exp \Big(-\frac{(r-r_0)^2}{2 D^2_T l}\Big),
		\end{eqnarray}
	 where $\phi_L$ and  $\phi_T$ is longitudinal and transverse gaussian distributions, $D_L$ and $D_T$ is longitudinal and transvers diffusion constants which are calculated using MAGBOLTZ \cite{BIAGI1989716}, $z_0$ and $r_0$ is the position of center of mass of the distribution, $l$ is the drifted distance at time t, r and z are the position of electron in cylindrical co-ordinate system. When the number of electrons and ions became of the order of $\approx 10^7$ the simulation is stopped and the position of electrons and ions are stored (see figures \ref{avalanche_electron_dist} and \ref{avalanche_ion_dist}) to calculate the electric field.  \\
	In the following subsections the calculation of space charge field
	along with image charge field has been done using three models,
	(a) charged ring, (b) line charge, and (c) neBEM and the comparison between them also been discussed.

	\subsection{Field of a uniformly charged ring}
	The electric field of the charge distribution shown in figures \ref{avalanche_electron_dist} and \ref{avalanche_ion_dist} can be calculated by modeling the charge region as a number of concentric rings. The field at any point $(r,\phi,z)$ due to a uniformly charge ring of radius $r^\prime$ located at $z^\prime$ can be expressed as \cite{Lippmann_1,LippmanThesis}, 
	
	\begin{subequations}
		\begin{align}
		E_{r}^{ring}(r,z,r^{\prime},z^{\prime})&\approx \frac{Q}{2\pi\epsilon_{0}}\frac{1}{ra^2b}\times
		\left[ c^{2}E\left(\frac{-4rr^{\prime}}{b^{2}}  \right) +a^{2}K\left(\frac{-4rr^{\prime}}{b^{2}}  \right)\right]\label{eqn:Er_lip}  \\
		E_{\phi}^{ring}(r,z,r^{\prime},z^{\prime})&=0\label{eqn:Ephi_lip} \\
		E_{z}^{ring}(r,z,r^{\prime},z^{\prime})&\approx\frac{Q}{\pi \epsilon_{0}}\frac{(z-z^\prime)}{a^2b}E\left(\frac{-4rr^{\prime}}{b^{2}}\right)\label{eqn:Ez_lip}
		\end{align}
	\end{subequations}
	where,
	
	\begin{subequations}
		\begin{align}
		a^2&=(r+r^\prime)^2+(z-z^\prime)^2\\
		b^2&=(r-r^\prime)^2+(z-z^\prime)^2\\
		c^2&=r-(r^\prime)^2+(z-z^\prime)^2
		\end{align}
	\end{subequations}
	and
	
	\begin{align}
	K(x)&=\int\limits_{0}^{\frac{\pi}{2}}\frac{1}{\sqrt{1-x \sin^2(\zeta)}}d\zeta\\
	E(x)&=\int\limits_{0}^{\frac{\pi}{2}}\sqrt{1-x \sin^2(\zeta)}\,\,d\zeta
	\end{align}
	$K(x)$ and $E(x)$ represents the first and second kind elliptic integrals respectively.	The steps of finding charge inside the rings are also discussed in \cite{Lippmann_1}. The field of  mirror charged rings located at $(r^\prime,\phi^\prime,2g-z^\prime)$ and $(r^\prime,\phi^\prime,-z^\prime)$ can be found by replacing $z^\prime$ with $2g-z^\prime$ and $-z^\prime$ in equations \ref{eqn:Er_lip} and \ref{eqn:Ez_lip}.
	
	\subsection{Field due to a single line charge}
		In the above ring approximation, due to the rotational symmetry of the avalanche charged region, the $\phi$ directional field $E_{\phi}$ is considered zero. However, depending on the experimental situation, the charge region may not be properly rotationally symmetric \cite{Dey_2020}. So the approximation of uniform ring is not always good enough. Hence, in those cases, one can divide the ring into several uniformly charged straight lines along the periphery where each can carry a different charge. Therefore, the sum of all lines over a ring together can be represented as a non-uniform charged ring. The method of division of rings in several lines is discussed in \cite{Dey_2020}.  
	The electric field at any position (x,y,z) due to a line of uniform charged density $\bar{\lambda}$ and length S, located at $x^\prime=\bar{r}$,$z^\prime=\bar{z}$ and parallel to y-axis can be expressed as follows \cite{Dey_2020},
	
	\begin{subequations}
		\begin{align}
		E_{x}^{line}&=\frac{\bar{\lambda}(x-\bar{r})}{4\pi\epsilon_{0}P^{2}}\left[\frac{(y+\frac{S}{2})}{\sqrt{(y+\frac{S}{2})^{2}+P^{2}}}-\frac{(y-\frac{S}{2})}{\sqrt{(y-\frac{S}{2})^{2}+P^{2}}}\right]\label{eqn:ex_line}\\
		E_{y}^{line}&=-\frac{\bar{\lambda}}{4\pi\epsilon_{0}}\left[\frac{1}{\sqrt{(y+\frac{S}{2})^{2}+P^{2}}}-\frac{1}{\sqrt{(y-\frac{S}{2})^{2}+P^{2}}}\right]\label{eqn:ey_line}\\
		E_{z}^{line}&=\frac{\bar{\lambda}(z-\bar{z})}{4\pi\epsilon_{0}P^{2}}\left[\frac{(y+\frac{S}{2})}{\sqrt{(y+\frac{S}{2})^{2}+P^{2}}}-\frac{(y-\frac{S}{2})}{\sqrt{(y-\frac{S}{2})^{2}+P^{2}}}\right]\label{eqn:ez_line}
		\end{align}
	\end{subequations}
	where $P=\sqrt{(z-\bar{z})^{2}+(x-\bar{r})^{2}}$, and if $Q_{st}$
	is the total charge of this straight line then, $\bar{\lambda}=\frac{Q_{st}}{S}$. The field of a mirror line charge at any point inside the gas gap g  can be found by replacing $\bar{z}$ with the position of the mirror line from the respective electrode in equations \ref{eqn:ex_line},\ref{eqn:ey_line} and \ref{eqn:ez_line}. The positions of mirror lines can be found from table \ref{tab:threelayer_image} and \ref{tab:genaration of image RPC}.
	
		
	\subsection{Comparison of Z-directional field $E_{z}$}
	\subsubsection{  Ring and line approximation}
	The combined z-directional field of avalanche electron and ion distribution (see figures \ref{avalanche_electron_dist} and \ref{avalanche_ion_dist}) has been calculated using equations \ref{eqn:Ez_lip} and \ref{eqn:ez_line}, where both radial and z-directional thickness of each ring is 0.001 cm. As discussed in the \cite{Dey_2020}, to segment a ring in several lines, one needs to assign one more parameter $\delta\phi$, taken as 1 degree in this calculation, where $\delta\phi$ is the angle, subtended to the center of the circular ring.  Here it is considered that the charge enclosed by a ring is uniformly distributed over the ring. Therefore after segmentation, each line will carry the same amount of charge, which is similar to case-1 in \cite{Dey_2020}. 
	\begin{figure}
		\center\subfloat[\label{fig:ezfieldlip}]{\includegraphics[scale=0.4]{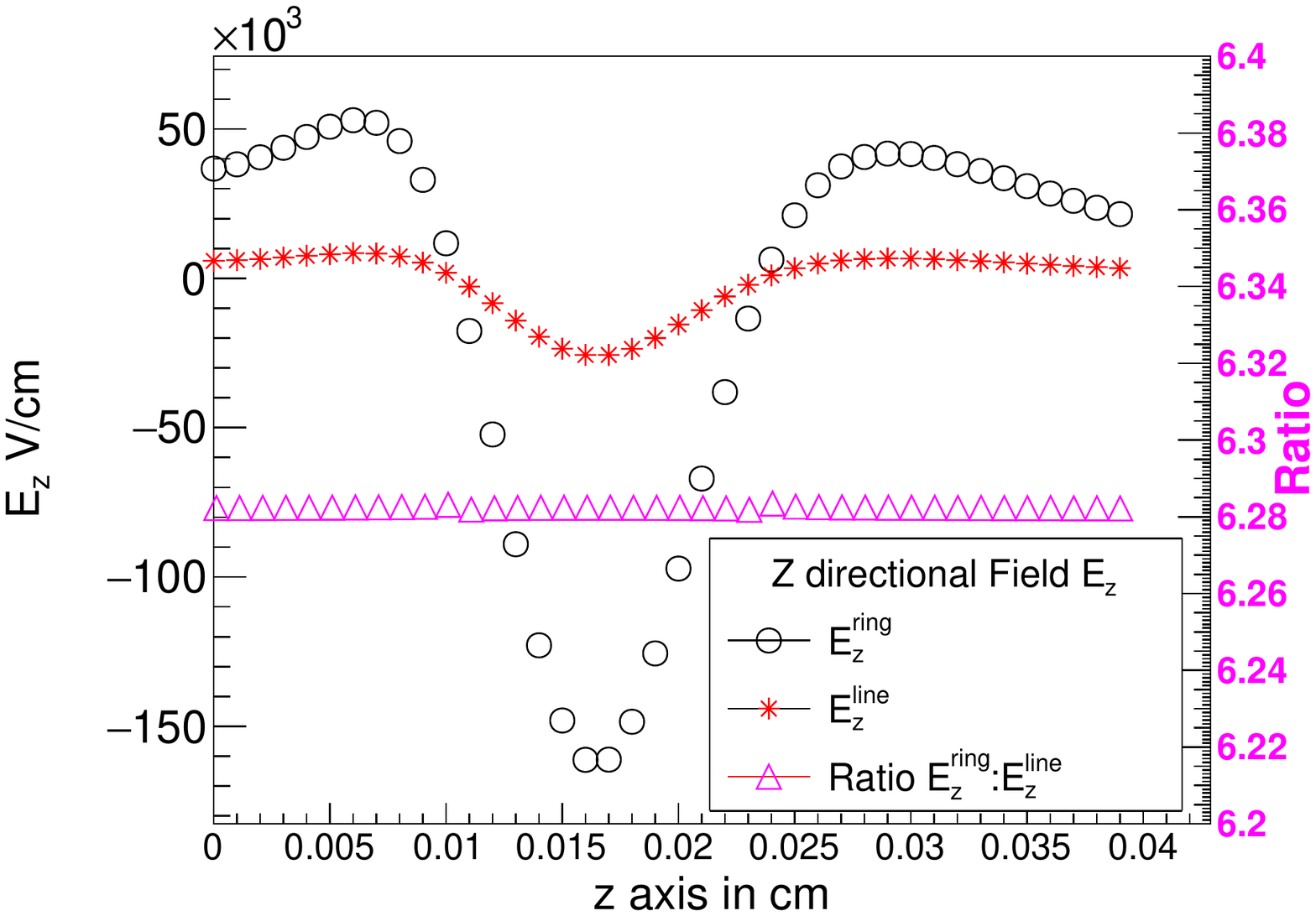}
			
		}\subfloat[\label{fig:ezfieldlipCorrected}]{\includegraphics[scale=0.4]{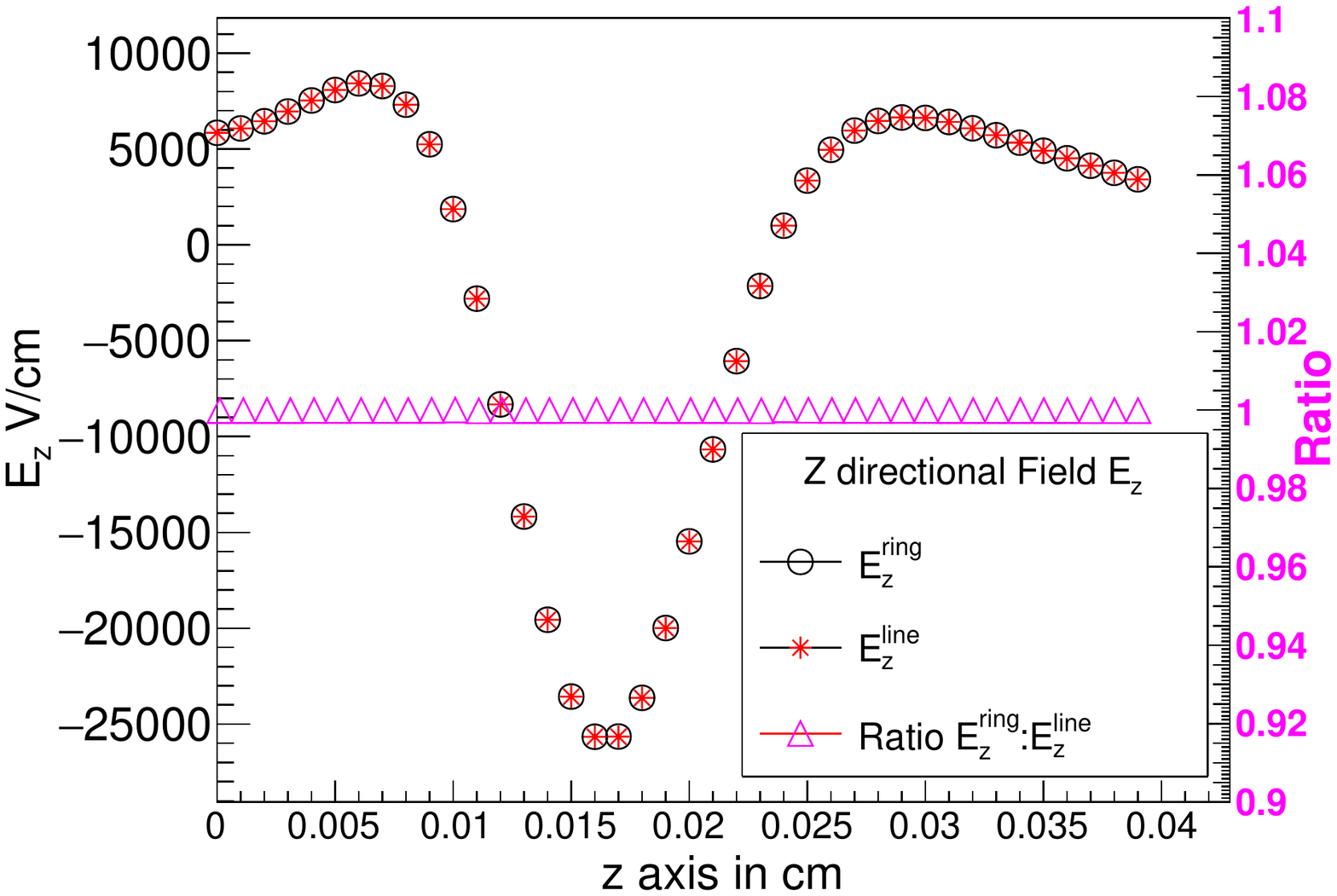}
			
		}
		
		\caption{(a) Comparison between ring and line z-directional field (source+image) before $2\pi$ division. (b) Comparison between ring and line z-directional field (source+image) after $2\pi$ division.}
	\end{figure}
	\begin{figure}
	\center\subfloat[ \label{fig:ezfieldImagelip}]{\includegraphics[scale=0.4]{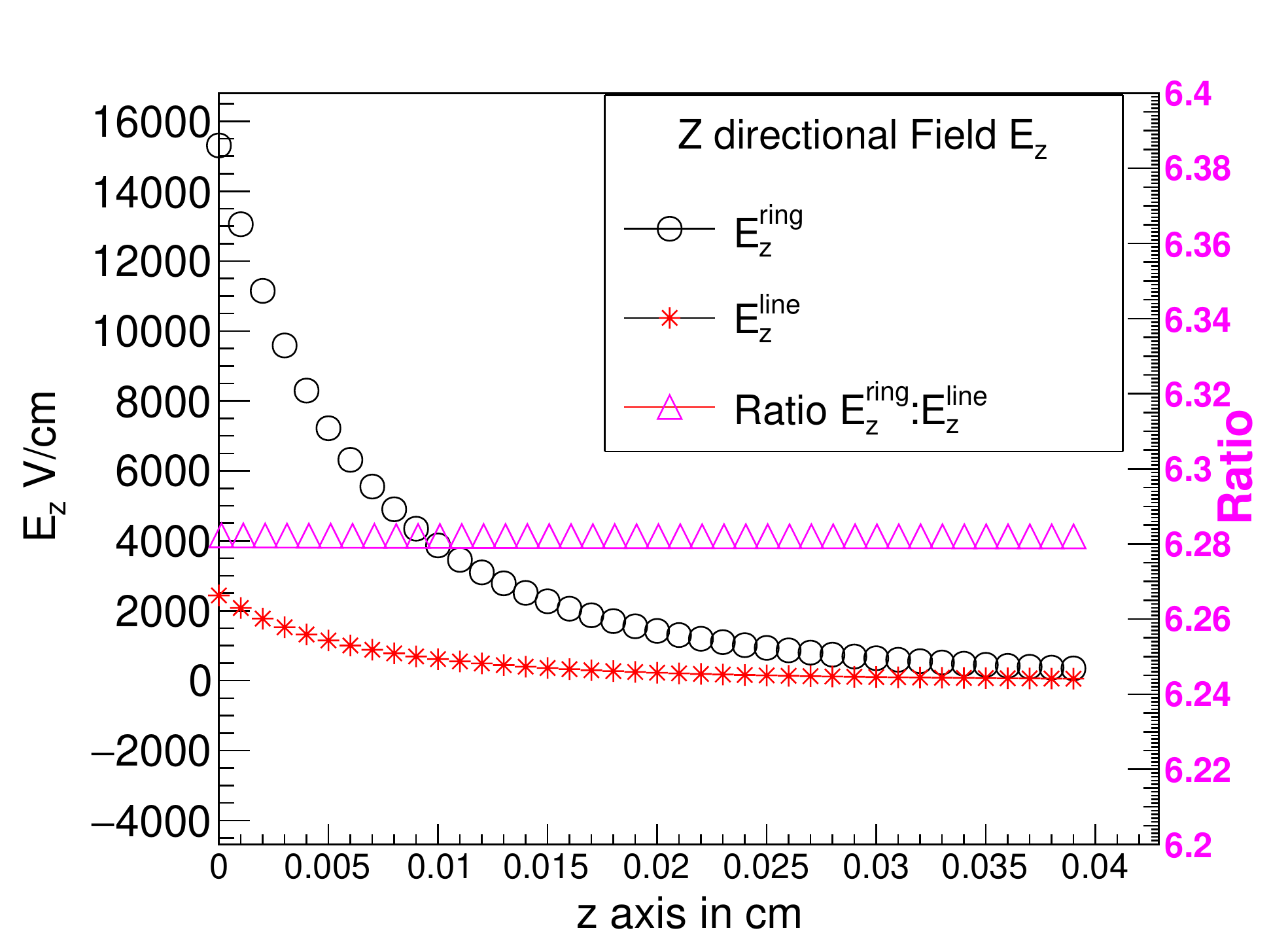}
	}\subfloat[ \label{fig:ezfieldImageLipCorrected}]{\includegraphics[scale=0.4]{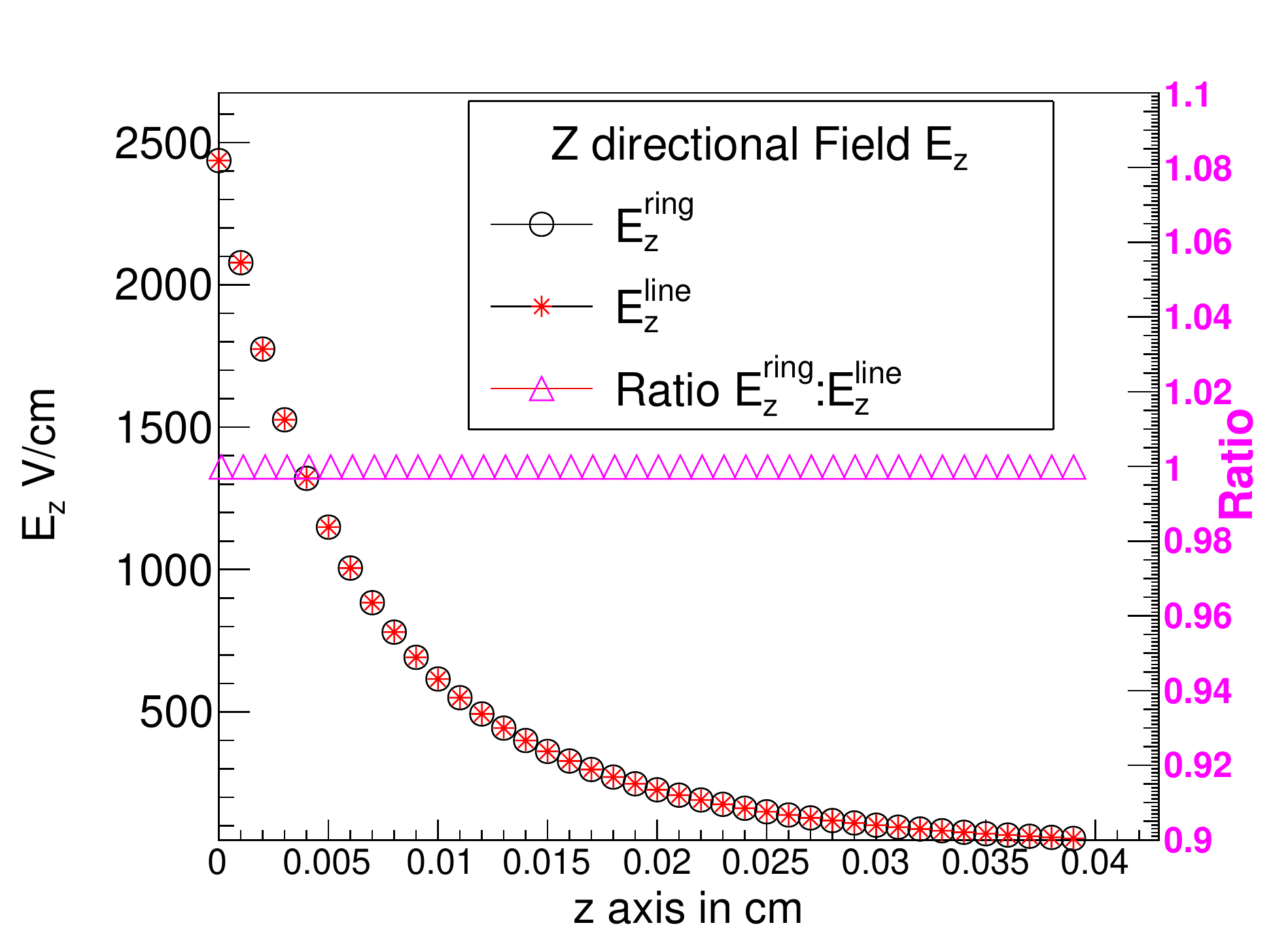}
		
	}
	
	\caption{(a)
		Comparision between ring and line z-directional image field before $2\pi$ division. (b) Comparision between ring and line z-directional image field after $2\pi$ division.}
\end{figure}
The variation of total (source+image) and only image z-directional field along the z-axis inside the gas gap for both ring and line approximation with their ratios at each point $(E_{z}^{ring}:E_{z}^{line})$ have been shown in figures \ref{fig:ezfieldlip} and \ref{fig:ezfieldImagelip} respectively. It is found that in both cases the ratio at each point is $\frac{E_{z}^{ring}}{E_{z}^{line}}\approx6.28\approx\,2\pi$ approximately. Therefore the field values are large for ring approximation in figures \ref{fig:ezfieldlip} and \ref{fig:ezfieldImagelip}. If we divide the values of $E_{z}^{ring}$ with $2\pi$ then the field values match exactly for both ring and line approximations, and the ratio becomes $\frac{E_{z}^{ring}}{E_{z}^{line}}\approx1$ (see figure \ref{fig:ezfieldlipCorrected} and \ref{fig:ezfieldImageLipCorrected}).
	\subsubsection{ neBEM and line approximation}
\begin{figure}
		\center\subfloat[\label{fig:neBEM_no_image} ]{\includegraphics[scale=0.4]{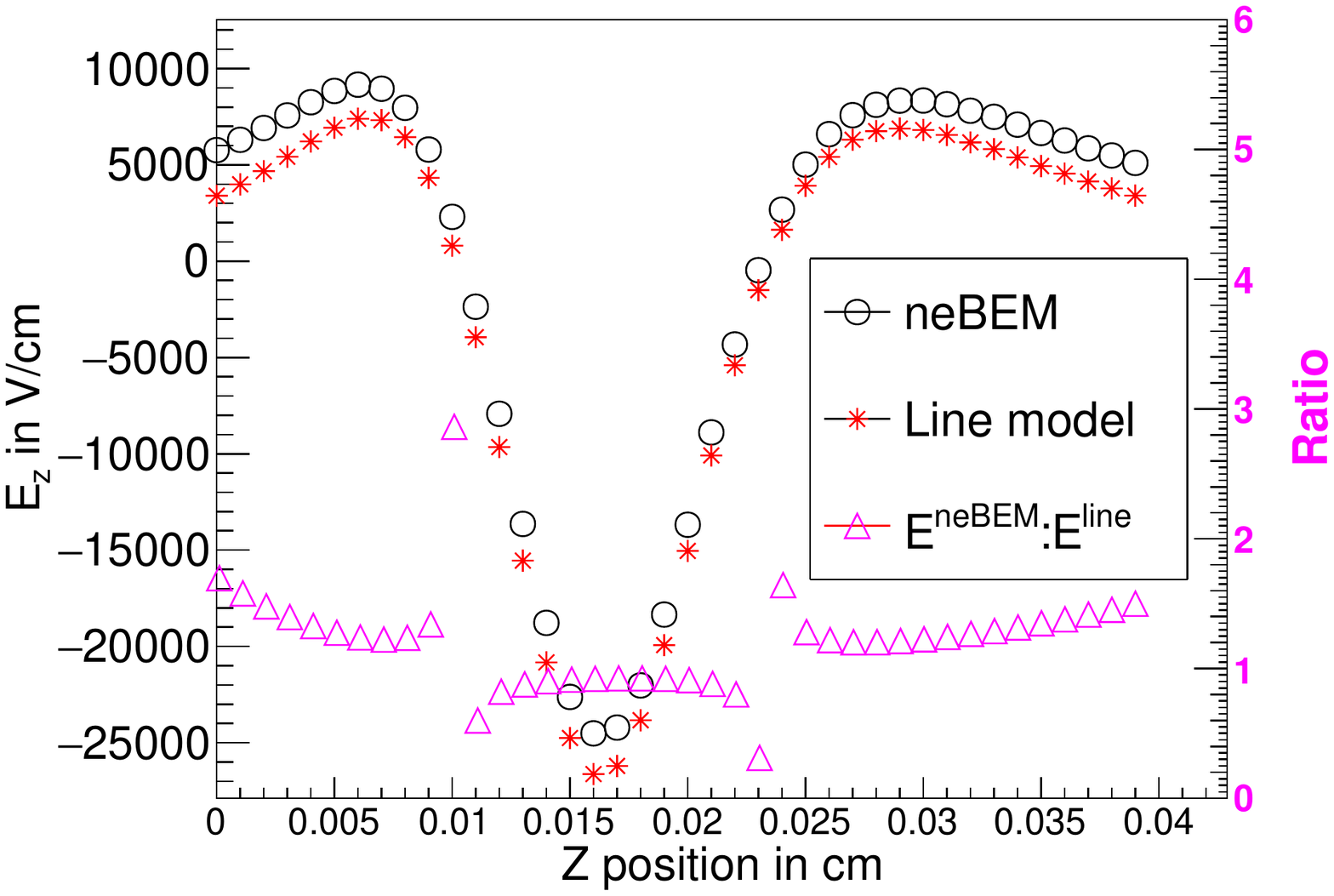}
			
		}\subfloat[\label{fig:neBEM_with_image} ]{\includegraphics[scale=0.4]{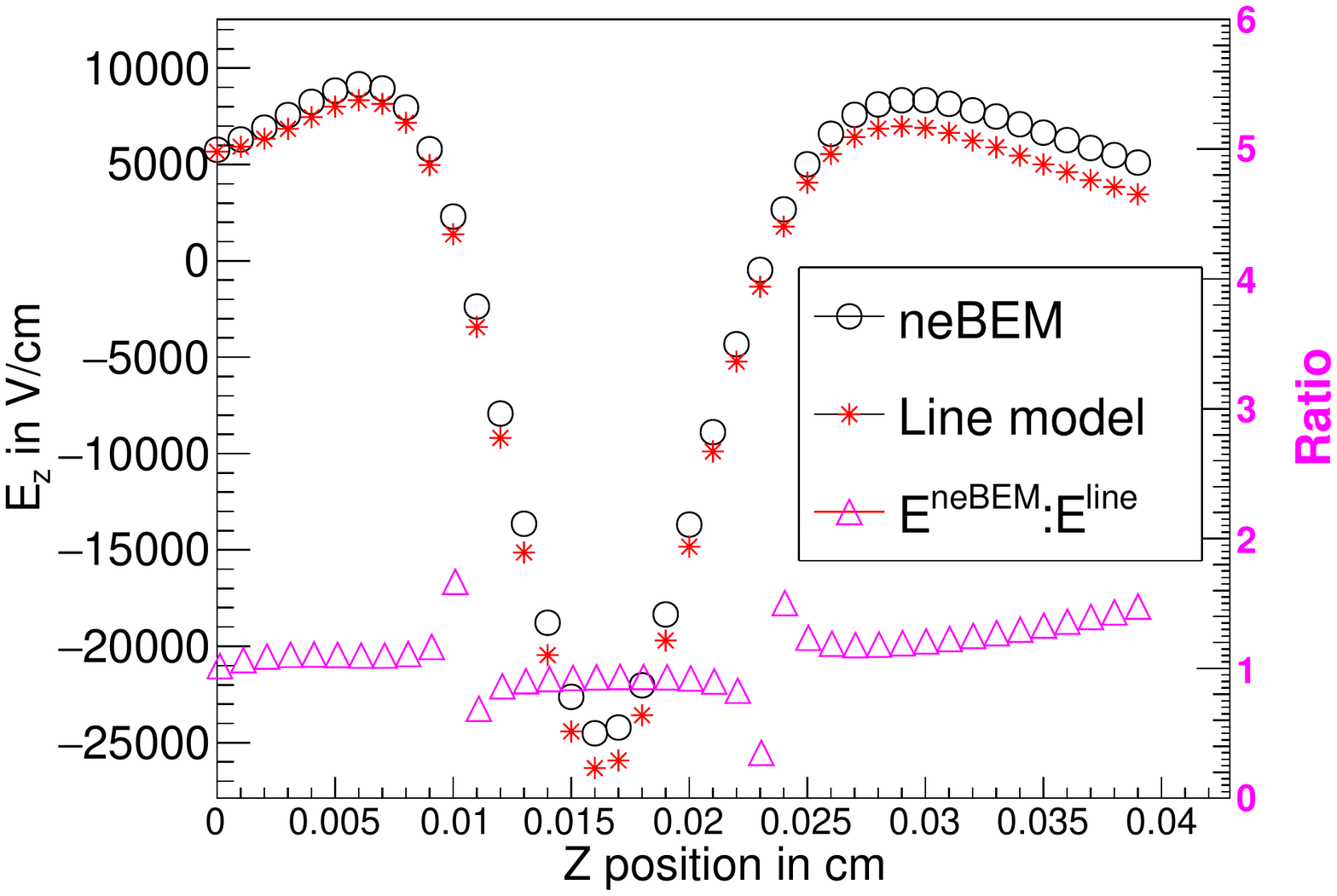}
			
		}
		
		\caption{(a) Comparision of  z-directional field without considering effects due to image charges with neBEM. (b) Comparision of total z-directional field (source+image) with neBEM. }
	\end{figure}

	In figures \ref{fig:neBEM_no_image} and \ref{fig:neBEM_with_image}, we have compared the results obtained using the proposed method with those obtained using the numerical solver neBEM. In the former figure, line model without image charge has been used to estimate the field. The estimated fields are different from those obtained using neBEM, the larger deviations (ratio between the estimates is $\approx
	1.75$) being close to the RPC surface where the effect of image charges are expected to be important. Beyond this region, the comparison is reasonable-the ratio between the estimated values varying from 0.8 to 1.25. The ratio is much larger close to regions where the field values themselves are close to zero and can be safely ignored. The estimates from the line model with image charge have been compared to the same neBEM estimates in the latter figure. Here, remarkable improvement is observed close to the RPC surface, the ratio between the two estimates being close to 1. In the rest of the domain, the quality of agreement remains unchanged. This comparison clearly shows the efficacy of the proposed approach to incorporate effects of image charge using the line model.    
	\vfill
	\subsection{Selection of number of image charge for avalanche charge distribution} \label{Subsec:Selection_image_avCharge}
		\begin{figure}
		\center\subfloat[ \label{fig:deltaSm_avalanche_chargeD1}]{\includegraphics[scale=0.38]{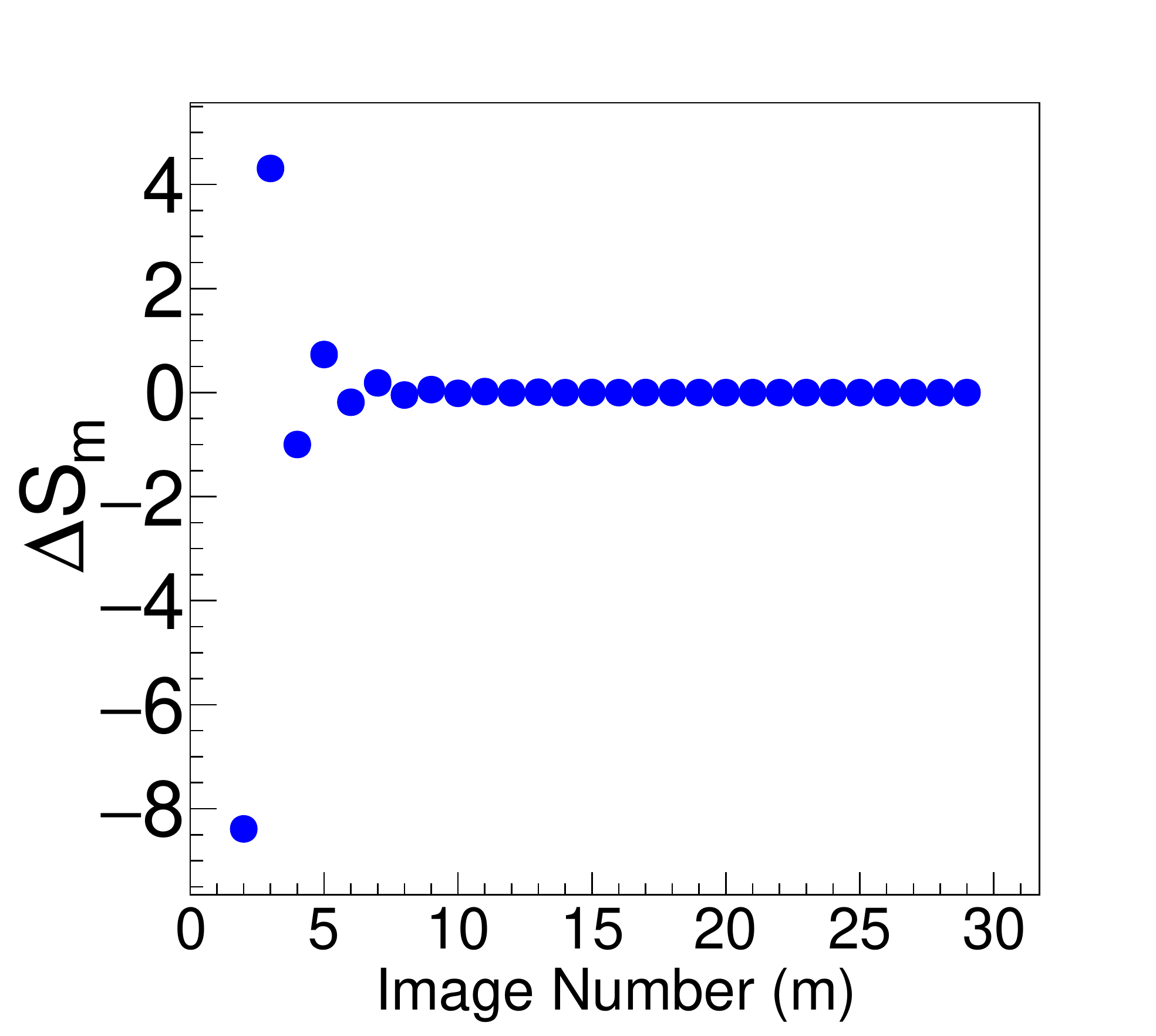}
			
		}\subfloat[ \label{fig:deltaSm_avalanche_chargeD2}]{\includegraphics[scale=0.38]{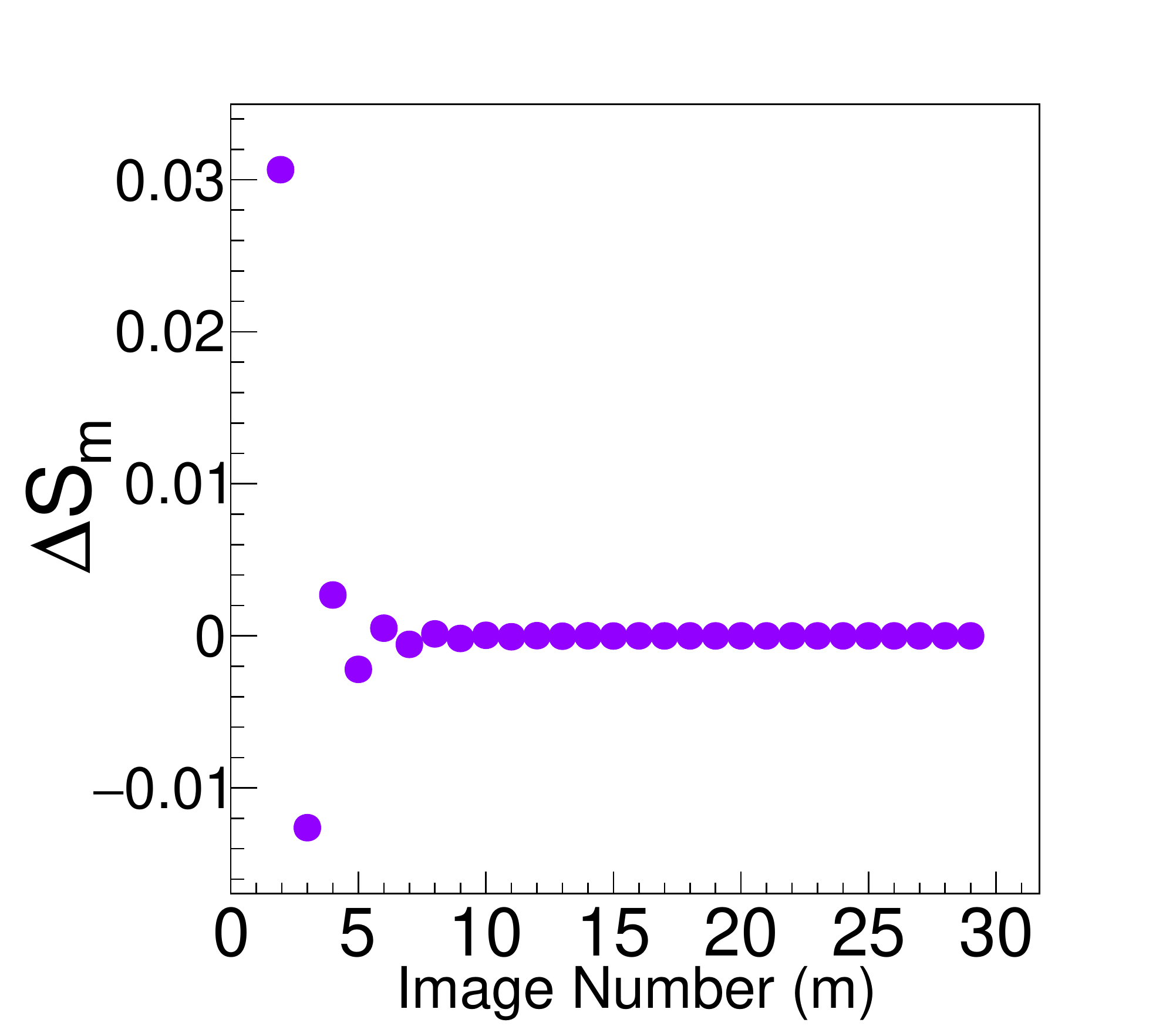}
			
		}
		
		\caption{ (a) Percentage of contribution of higher order avalanche image charges on the total electric field
			for dielectric electrode D1 of figure \ref{fig:RPC_image_charge_formation}. (b) Percentage of contribution of higher order avalanche image charges on the total electric
			field for dielectric electrode D2 of figure \ref{fig:RPC_image_charge_formation}.
		}
	\end{figure}

	In subsection (\ref{subsec:6.3}), we have discussed the selection criteria of the number of images due to a single point charge inside RPC. This section will discuss the same for the avalanche charge distribution. The value of $\Delta S_{m}$ for both electrodes D1 and D2 has been shown in figures \ref{fig:deltaSm_avalanche_chargeD1} and  \ref{fig:deltaSm_avalanche_chargeD2}, where the results are more or less similar with single point charge case. Since the charge cluster is near electrode D2, the first-order term $Q_{31}^\prime$ of $M5^{D2}$ has more contribution on the field than the other higher-order terms. Hence, the maximum contribution of higher-order terms or the absolute maximum value of  $\Delta S_{m}$ ($\approx0.03\%$) is minimal.
	On the other hand, electrode D1 is far from the charge cluster. Hence, the first-order term $Q_{31}$ of $M5^{D1}$ is itself a very weak contributor, and so are the other corresponding higher orders. Therefore, the maximum absolute value of $\Delta S_{m}$ ( $\approx 8\%$) for D1 is found to be larger than maximum absolute value $\Delta S_{m}$ for D2 (m=2,3,4..). However, the strength of the field is significantly less for the images of electrode D1.	
	
	\subsection{Variation of image field with electrode material}\label{sec:varImFieldWPer}
	
\begin{figure}
		\center{\includegraphics[scale=0.4]{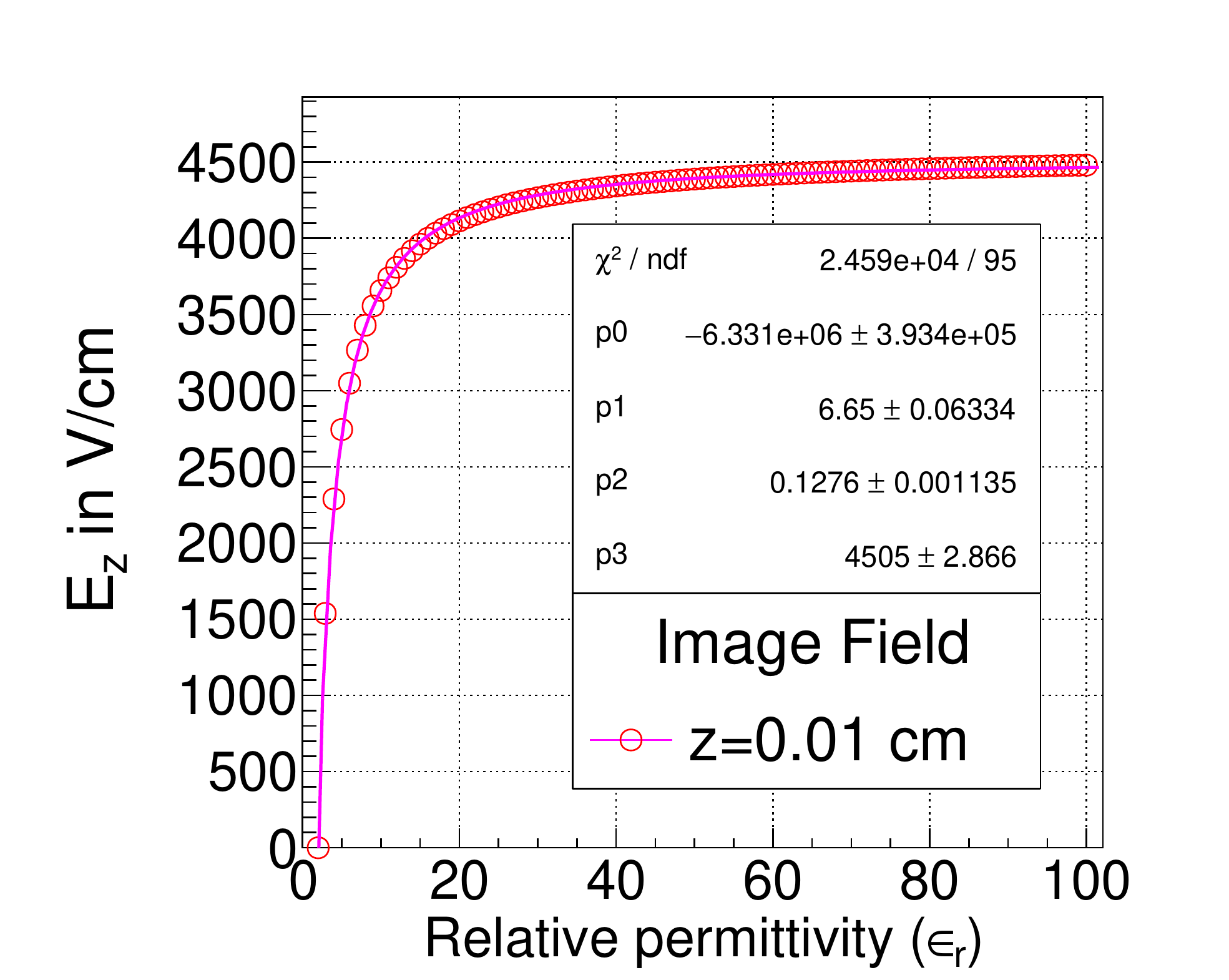}
			
		}
		
		\caption{\label{fig:Ez_vs_epr_image}Variation of avalanche image field at z=0.01 cm with relative permittivity of electrodes ($\epsilon_{r}$) of an RPC }
	\end{figure}

	The variation of the applied electric field with the relative permittivity of the electrode has been discussed in section \ref{sec:section2} (case 1). We have seen in sections \ref{section5} and \ref{sec:sec_6} that the contribution of the higher-order image charges is very small, and from the table \ref{tab:threelayer_image} and \ref{tab:genaration of image RPC} it is confirmed that the position of image charges depends on the electrode thickness and gas-gap. Therefore, it can be concluded that the dependence of the image charge field on the electrode thickness and gas gap is also less. Thus, we will only discuss the variation of the image field on $\epsilon_{r}$. As higher-order terms are less significant, we neglect them from $M5^{D1, D2}$, and then we are left with only a single image charge $Q\,\alpha_{32}$ for D1 and $Q\,\alpha_{34}$ for D2. Therefore,  the image field at any point (x,y,z) inside the gas gap can be expressed as:
	
	\begin{align}
	E^{image}_{z}=E_{z}^{D1}(Q\,\alpha_{32},h,g,x,y,z,\bar{r},\bar{z})+E_{z}^{D2}(Q\,\alpha_{34},h^\prime,g,x,y,z,,\bar{r},\bar{z}).
	\end{align}
	Here we are using line equation \ref{eqn:ez_line} to calculate both $E_z^{D1,D2}$. As $\alpha_{32}=\alpha_{34}$ we can write,  
	   
	   \begin{equation}
	   \begin{split}
	   E^{image}_{z}&=\alpha_{32}(E_{z}^{D1}(Q,h,g,x,y,z,,\bar{r},\bar{z})+E_{z}^{D2}(Q,h^\prime,g,x,y,z,,\bar{r},\bar{z})).
	   \end{split}
	   \end{equation}
	   From equation \ref{eqn:alpha32} we can write, $\alpha_{32}=\frac{1-\epsilon_{r}}{1+\epsilon_{r}
	   }$, where  $\epsilon_{r}$=$\frac{\epsilon_{2}}{\epsilon_{0}}$ and $\epsilon_{0}$=permittivity of gas/air. The variation of $E_z^{image}$ with relative permittivity $\epsilon_{r}$ of electrode has been shown in figure \ref{fig:Ez_vs_epr_image} at z=0.01cm, where it is found that on an increment of $\epsilon_{r}$, the field value also increases first, then it starts showing the saturation after $\epsilon_{r}>20$. The data points of figure \ref{fig:Ez_vs_epr_image} are fitted with the equation:
    \begin{equation}\label{eqn:4.8}
    f(\epsilon_{r})=p0\,exp(-p1\,\epsilon_{r}^{p2})+p3.
    \end{equation}
    The fitted values of the parameters p0,p1,p2,p3 have been shown in the same figure.
    Again like equation \ref{eqn:2.1} the parameters p0 and p3 of equation \ref{eqn:4.8} have the dimension of the electric field, and p1 is the function of $\epsilon_r$ and p2 must be constant. Other physical significance of the same parameters are yet to be understood, but the functional form allows interpolation for arbitrary values of $\epsilon_{r}$.
	
	\section{Summary}
	 We have discussed the dependence of the applied electric field inside an RPC with electrode parameters such as permittivity, the thickness of electrodes, and the gas gap. It is seen that at fixed applied voltage, electrode thickness, and gas gap, the electric field increases rapidly with permittivity and starts to saturate approximately after the value of 20. Also, the variation of the electric field with the permittivity of the electrode decreases for the smaller thickness of the electrodes.  
	 On the other hand, the electric field reduces with the increment of electrode thickness when the permittivity and gas gap is fixed. The electric field inside the middle of the gas gap diminishes on the increment of the gas gap, as expected.
	 \par In sections \ref{sec:section3} and \ref{sec:Section4_avalancheCharge}, the calculation of the total field (source + image) of charges at arbitrary locations in an RPC has been discussed. The electric field caused by the space charge-induced dipoles on the electrodes has been calculated using the method of images. This enables us to compute the total field-induced due to the presence of an avalanche. We have also given an example of electric field calculation of an avalanche charge distribution where we have used a straight-line model. The validity of our straight-line model with existing models in the literature has also been checked, and a very good agreement is observed. 
	 \par It is clear that the number of space charges of a growing avalanche is a stochastic process, and the image charge field depends on the number of space charges and their distances from the electrodes. Therefore, at every step of a growing avalanche, the number and distance of space charges and the images of them will change. Hence,  we have proposed a technique to dynamically choose reflections of the image charges while the avalanche is growing.   
	 Another important issue is the variation of the image field with permittivity of the electrodes, which is discussed in subsection \ref{sec:varImFieldWPer}. The image field is increased with permittivity, but approximately after $\epsilon_{r}>20$, the field value attains saturation.
	 \par The advantage of using the line model is that it removes the constraint of rotational symmetry of the avalanche charge region. Since the electric field equations are analytical and do not include any numerical integration, the method is fast and valuable while simulating an avalanche or avalanche to streamer transition. In the case of streamers, the charge can be distributed over the entire gas gap. Hence, the space charge electric field and image field variation will need to be further investigated. We hope to incorporate the proposed model in Garfield++ in the near future.
	 \section*{Acknowledgement}
	 The author Tanay Dey is grateful to the INO collaboration and the HEP experiment division of VECC for providing the resources and help. Also, we are thankful to the reviewer for helping us improve the content of the paper.

	\bibliographystyle{JHEP.bst}
	\bibliography{Image_issues_inRPC.bib}

\providecommand{\href}[2]{#2}\begingroup\raggedright\begin{thebibliography}{10}

\bibitem{cardeli-1}
R.~Santonico and R.~Cardarelli, \emph{Development of resistive plate counters},
  \href{http://dx.doi.org/https://doi.org/10.1016/0029-554X(81)90363-3}{\emph{Nuclear
  Instruments and Methods in Physics Research} {\bfseries 187} (1981)
  377--380}.

\bibitem{cardeli-2}
R.~Cardarelli, R.~Santonico, A.~Biagio and A.~Lucci, \emph{Progress in
  resistive plate counters},
  \href{http://dx.doi.org/https://doi.org/10.1016/0168-9002(88)91011-X}{\emph{Nuclear
  Instruments and Methods in Physics Research Section A: Accelerators,
  Spectrometers, Detectors and Associated Equipment} {\bfseries 263} (1988) 20
  -- 25}.

\bibitem{Goswami_2017}
S.~Goswami, \emph{The status of {INO}},
  \href{http://dx.doi.org/10.1088/1742-6596/888/1/012025}{\emph{Journal of
  Physics: Conference Series} {\bfseries 888} (sep, 2017) 012025}.

\bibitem{Kumari_2020}
P.~Kumari, K.~Lee, A.~Gelmi, K.~Shchablo, A.~Samalan, M.~Tytgat et~al.,
  \emph{Improved-{RPC} for the {CMS} muon system upgrade for the {HL}-{LHC}},
  \href{http://dx.doi.org/10.1088/1748-0221/15/11/c11012}{\emph{Journal of
  Instrumentation} {\bfseries 15} (nov, 2020) C11012--C11012}.

\bibitem{Collaboration_2012}
A.~Collaboration, F.~Boss{\`{u}}, M.~Gagliardi and M.~Marchisone,
  \emph{Performance of the {RPC}-based {ALICE} muon trigger system at the
  {LHC}}, \href{http://dx.doi.org/10.1088/1748-0221/7/12/t12002}{\emph{Journal
  of Instrumentation} {\bfseries 7} (dec, 2012) T12002--T12002}.

\bibitem{MONDAL2021166042}
M.~Mondal, T.~Dey, S.~Chattopadhyay, J.~Saini and Z.~Ahammed, \emph{Performance
  of a prototype bakelite rpc at gif++ using self-triggered electronics for the
  cbm experiment at fair},
  \href{http://dx.doi.org/https://doi.org/10.1016/j.nima.2021.166042}{\emph{Nuclear
  Instruments and Methods in Physics Research Section A: Accelerators,
  Spectrometers, Detectors and Associated Equipment} (2021) 166042}.

\bibitem{Cvetanovi__2015}
N.~Cvetanovi{\'{c}}, M.~M. Martinovi{\'{c}}, B.~M. Obradovi{\'{c}} and M.~M.
  Kuraica, \emph{Electric field measurement in gas discharges using stark
  shifts of he i lines and their forbidden counterparts},
  \href{http://dx.doi.org/10.1088/0022-3727/48/20/205201}{\emph{Journal of
  Physics D: Applied Physics} {\bfseries 48} (apr, 2015) 205201}.

\bibitem{AMMOSOV1997217}
V.~Ammosov, V.~Korablev and V.~Zaets, \emph{Electric field and currents in
  resistive plate chambers},
  \href{http://dx.doi.org/https://doi.org/10.1016/S0168-9002(97)00800-0}{\emph{Nuclear
  Instruments and Methods in Physics Research Section A: Accelerators,
  Spectrometers, Detectors and Associated Equipment} {\bfseries 401} (1997)
  217--228}.

\bibitem{MAJUMDAR2008346}
N.~Majumdar, S.~Mukhopadhyay and S.~Bhattacharya, \emph{Computation of 3d
  electrostatic weighting field in resistive plate chambers},
  \href{http://dx.doi.org/https://doi.org/10.1016/j.nima.2008.07.033}{\emph{Nuclear
  Instruments and Methods in Physics Research Section A: Accelerators,
  Spectrometers, Detectors and Associated Equipment} {\bfseries 595} (2008)
  346--352}.

\bibitem{MAJUMDAR2009719}
N.~Majumdar, S.~Mukhopadhyay and S.~Bhattacharya, \emph{Three-dimensional
  electrostatic field simulation of a resistive plate chamber},
  \href{http://dx.doi.org/https://doi.org/10.1016/j.nima.2008.12.098}{\emph{Nuclear
  Instruments and Methods in Physics Research Section A: Accelerators,
  Spectrometers, Detectors and Associated Equipment} {\bfseries 602} (2009)
  719--722}.

\bibitem{comsol}
{\emph{\href{https://www.comsol.co.in/}{https://www.comsol.co.in/}} }.

\bibitem{MOSHAII2012S168}
A.~Moshaii, L.~{Khosravi Khorashad}, M.~Eskandari and S.~Hosseini, \emph{Rpc
  simulation in avalanche and streamer modes using transport equations for
  electrons andions},
  \href{http://dx.doi.org/https://doi.org/10.1016/j.nima.2010.09.133}{\emph{Nuclear
  Instruments and Methods in Physics Research Section A: Accelerators,
  Spectrometers, Detectors and Associated Equipment} {\bfseries 661} (2012)
  S168--S171}.

\bibitem{Cardarelli1996AvalancheAS}
R.~Cardarelli, V.~Makeev and R.~Santonico, \emph{Avalanche and streamer mode
  operation of resistive plate chambers}, {\emph{Nuclear Instruments \& Methods
  in Physics Research Section A-accelerators Spectrometers Detectors and
  Associated Equipment} {\bfseries 382} (1996) 470--474}.

\bibitem{Lippmann_1}
C.~Lippmann and W.~Riegler, \emph{{Space charge effects in resistive plate
  chambers}}, \href{http://dx.doi.org/10.1016/j.nima.2003.08.174}{\emph{Nucl.
  Instrum. Meth. A} {\bfseries 517} (2004) 54--76}.

\bibitem{rpc-book}
P.~F. Marcello~Abbrescia, Vladimir~Peskov, \emph{{Resistive Gaseous Detectors:
  Designs, Performance, and Perspectives}}.
\newblock WILEY-VCH, (2018).

\bibitem{Paolozzi:2012pt}
L.~Paolozzi, G.~Aielli, R.~Cardarelli, A.~Di~Ciaccio, L.~Di~Stante, B.~Liberti
  et~al., \emph{{Test for upgrading the RPCs at very high counting rate}},
  \href{http://dx.doi.org/10.22323/1.159.0065}{\emph{PoS} {\bfseries RPC2012}
  (2012) 065}.

\bibitem{Gonzalez-Diaz:2006qno}
D.~Gonzalez-Diaz, P.~Fonte, J.~A. Garzon and A.~Mangiarotti, \emph{{An
  analytical description of rate effects in timing RPCs}},
  \href{http://dx.doi.org/10.1016/j.nuclphysbps.2006.07.026}{\emph{Nucl. Phys.
  B Proc. Suppl.} {\bfseries 158} (2006) 111--117}.

\bibitem{ABBRESCIA20047}
M.~Abbrescia, \emph{The dynamic behaviour of resistive plate chambers},
  \href{http://dx.doi.org/https://doi.org/10.1016/j.nima.2004.06.119}{\emph{Nuclear
  Instruments and Methods in Physics Research Section A: Accelerators,
  Spectrometers, Detectors and Associated Equipment} {\bfseries 533} (2004)
  7--10}.

\bibitem{Aielli_2016}
G.~Aielli, P.~Camarri, R.~Cardarelli, A.~D. Ciaccio, L.~D. Stante, R.~Iuppa
  et~al., \emph{Improving the {RPC} rate capability},
  \href{http://dx.doi.org/10.1088/1748-0221/11/07/p07014}{\emph{Journal of
  Instrumentation} {\bfseries 11} (jul, 2016) P07014--P07014}.

\bibitem{Carboni:2003my}
G.~Carboni, S.~De~Capua, D.~Domenici, G.~Ganis, R.~Messi, E.~Santovetti et~al.,
  \emph{{A Model for RPC detectors operating at high rate}},
  \href{http://dx.doi.org/10.1016/S0168-9002(02)02082-X}{\emph{Nucl. Instrum.
  Meth. A} {\bfseries 498} (2003) 135--142}.

\bibitem{HEUBRANDTNER}
T.~Heubrandtner, B.~Schnizer, C.~Lippmann and W.~Riegler, \emph{Static electric
  fields in an infinite plane condenser with one or three homogeneous layers},
  \href{http://dx.doi.org/https://doi.org/10.1016/S0168-9002(02)00805-7}{\emph{Nuclear
  Instruments and Methods in Physics Research Section A: Accelerators,
  Spectrometers, Detectors and Associated Equipment} {\bfseries 489} (2002)
  439--443}.

\bibitem{Lippmann:2003ar}
C.~Lippmann, W.~Riegler and B.~Schnizer, \emph{{Space charge effects and
  induced signals in resistive plate chambers}},
  \href{http://dx.doi.org/10.1016/S0168-9002(03)01270-1}{\emph{Nucl. Instrum.
  Meth. A} {\bfseries 508} (2003) 19--22}.

\bibitem{Weber-W}
W.~E, \emph{{Electromagnetic Theory}}.
\newblock New York,dover,(1965), pg. 218-33.

\bibitem{Jomaa1983ElectricFD}
B.~M. Jomaa, \emph{Electric field distribution through multilayer environment
  in the vicinity of high voltage transmission lines},  1983.

\bibitem{multi-layer}
T.~Takashima and R.~Ishibashi, \emph{Electric fields in dielectric multi-layers
  calculated by digital computer},
  \href{http://dx.doi.org/10.1109/TEI.1978.298097}{\emph{IEEE Transactions on
  Electrical Insulation} {\bfseries EI-13} (1978) 37--44}.

\bibitem{Dey_2020}
T.~Dey, S.~Mukhopadhyay, S.~Chattopadhyay and J.~Sadukhan, \emph{Numerical
  study of space charge electric field inside resistive plate chamber},
  \href{http://dx.doi.org/10.1088/1748-0221/15/11/c11005}{\emph{Journal of
  Instrumentation} {\bfseries 15} (nov, 2020) C11005--C11005}.

\bibitem{new_comb_image_plate}
W.~A. Newcomb, \emph{Trouble with the method of images},
  \href{http://dx.doi.org/10.1119/1.12786}{\emph{American Journal of Physics}
  {\bfseries 50} (1982) 601--607},
  [\href{https://arxiv.org/abs/https://doi.org/10.1119/1.12786}{{\ttfamily
  https://doi.org/10.1119/1.12786}}].

\bibitem{BIAGI1989716}
S.~Biagi, \emph{A multiterm boltzmann analysis of drift velocity, diffusion,
  gain and magnetic-field effects in argon-methane-water-vapour mixtures},
  \href{http://dx.doi.org/https://doi.org/10.1016/0168-9002(89)91446-0}{\emph{Nuclear
  Instruments and Methods in Physics Research Section A: Accelerators,
  Spectrometers, Detectors and Associated Equipment} {\bfseries 283} (1989)
  716--722}.

\bibitem{LippmanThesis}
C.~Lippmann, \emph{Detector physics of resistive plate chambers
  (cern-thesis-2003-035)},  2003.

\end{thebibliography}\endgroup
	\newpage
	\appendix
	\section{Algorithm}\label{appendix}
	\begin{figure}[H]
	\center{\includegraphics[scale=0.45]{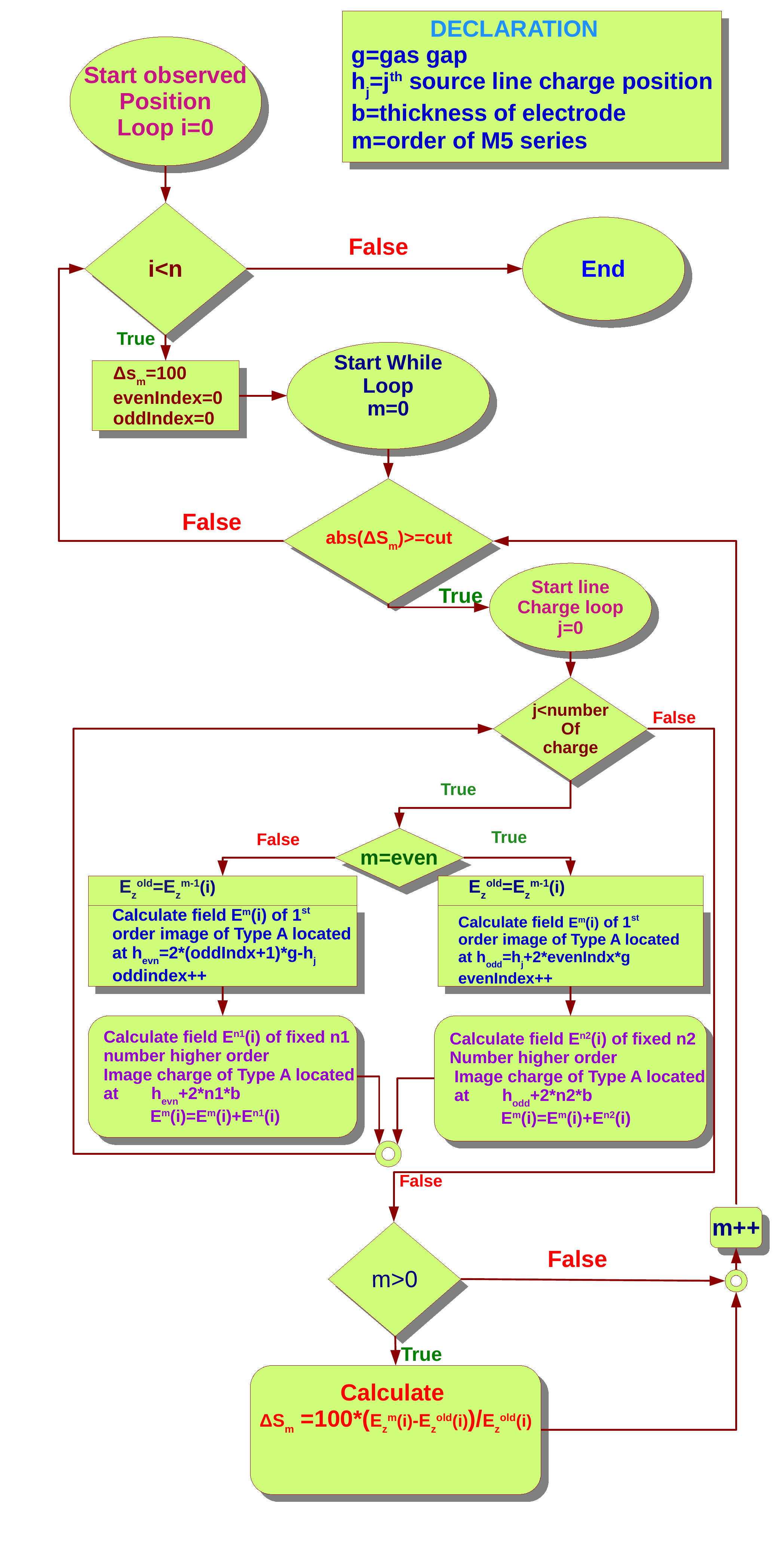}
		
	}
	
\end{figure}
	
\end{document}